\documentclass{aa}  

\usepackage{graphicx}
\usepackage{natbib}
\usepackage{multirow}
\usepackage{upgreek}
\usepackage[para,online,flushleft]{threeparttable}
\usepackage{txfonts}
\usepackage{hyperref}
\hypersetup{
  colorlinks,
  citecolor=blue,
  linkcolor=blue,
  urlcolor=blue}
\bibpunct{(}{)}{;}{a}{}{,} 
\newcommand{\psr}{PSR~J1933$-$6211 }
\newcommand{\psrc}{PSR~J1933$-$6211, }
\newcommand{\psrf}{PSR~J1933$-$6211.}
\newcommand{\omdot}{$\dot \omega$ }
\newcommand{\pbdot}{$\dot P_b$ }
\newcommand{\xdot}{$\dot x$ }
\newcommand{\msun}{\rm M_{\odot} }
\newcommand{\stig}{\varsigma}

\newcommand{\Mp}{M_{\rm p}}
\newcommand{\Mc}{M_{\rm c}}

\begin{document} 

   \title{Mass measurements and 3D orbital geometry of PSR J1933$-$6211}
    \author{M.~Geyer \inst{\ref{sarao}, \ref{uct}, \ref{mpifr}\dagger}
           \and V.~Venkatraman~Krishnan \inst{\ref{mpifr} \star}
           \and P.~C.~C.~Freire \inst{\ref{mpifr}}
           \and M.~Kramer \inst{\ref{mpifr}}
           \and J.~Antoniadis \inst{\ref{forth}, \ref{mpifr}}
           \and M.~Bailes\inst{\ref{swinburne}, \ref{ozgrav}}
           \and M.~C.~i~Bernadich\inst{\ref{mpifr}}
           \and S.~Buchner\inst{\ref{sarao}}
           \and A.~D.~Cameron \inst{\ref{swinburne}, \ref{ozgrav}} 
           \and D.~J.~Champion \inst{\ref{mpifr}}
           \and A.~Karastergiou \inst{\ref{oxford}, \ref{rhodes}}
           \and M.~J. Keith \inst{\ref{jbca}}
           \and M.~E. Lower \inst{\ref{atnf}}
           \and S.~Os{\l}owski\inst{\ref{manly}}
           \and A.~Possenti \inst{\ref{inaf}}
           \and A.~Parthasarathy \inst{\ref{mpifr}}
           \and D.~J.~Reardon \inst{\ref{swinburne}, \ref{ozgrav}}
           \and M.~Serylak \inst{\ref{skao}, \ref{uwc}}
           \and R.~M. Shannon \inst{\ref{swinburne}, \ref{ozgrav}}
           \and R.~Spiewak \inst{\ref{jbca}, \ref{swinburne}, \ref{ozgrav}}
           \and W.~van~Straten \inst{\ref{aut}}
           \and J.~P.~W.~Verbiest \inst{\ref{biele}, \ref{mpifr}}
          }

  \institute{South African Radio Astronomy Observatory, 2 
              Fir Street, Black River Park, Observatory 7925, South Africa\label{sarao}
              \and
              Department of Astronomy, University of Cape Town, Rondebosch, Cape Town, 7700, South Africa\label{uct}
              \and
              Max-Planck-Institut f\"{u}r Radioastronomie, Auf dem H\"{u}gel 69, D-53121 Bonn, Germany\label{mpifr}
              \and
              Institute of Astrophysics, FORTH, Dept. of Physics, University Campus, GR-71003 Heraklion, Greece \label{forth}
              \and
              Centre for Astrophysics and Supercomputing, Swinburne University of Technology, Hawthorn, VIC, 3122, Australia\label{swinburne}
              \and
              ARC Centre of Excellence for Gravitational Wave Discovery (OzGrav) \label{ozgrav}
              \and
              Department of Astrophysics, University of Oxford, Denys Wilkinson Building, Keble Road, Oxford OX1 3RH, UK\label{oxford}
              \and
              Department of Physics and Electronics, Rhodes University, PO Box 94, Grahamstown 6140, South Africa\label{rhodes}
              \and
              Jodrell Bank Centre for Astrophysics, Department of Physics and Astronomy, The University of Manchester, Manchester M13 9PL, UK \label{jbca} \and
              Australia Telescope National Facility, CSIRO, Space and Astronomy, PO Box 76, Epping, NSW 1710, Australia\label{atnf}
              \and 
              Manly Astrophysics, 15/41-42 East Esplanade, Manly 2095, NSW, Australia\label{manly} 
              \and 
              INAF-Osservatorio Astronomico di Cagliari, via della Scienza 5, I-09047 Selargius, Italy\label{inaf} 
              \and
              SKA Observatory, Jodrell Bank, Lower Withington, Macclesfield, SK11 9FT, United Kingdom\label{skao}
              \and
              Department of Physics and Astronomy, University of the Western Cape, Bellville, Cape Town, 7535, South Africa\label{uwc}
              \and
              Institute for Radio Astronomy \& Space Research, Auckland University of Technology, Private Bag 92006, Auckland 1142, NZ\label{aut}
              \and
              Fakultät für Physik, Universität Bielefeld, Postfach 100131, 33501, Bielefeld, Germany\label{biele}
              \\
              \email{$\dagger$marisa.geyer@uct.ac.za, $\star$vkrishnan@mpifr-bonn.mpg.de}
             }

   \date{Accepted to Astronomy \& Astrophysics -- 12 April 2023}

\abstract{\psr is a pulsar with a spin period of 3.5 ms in a 12.8 d nearly circular orbit with a white dwarf companion. Its high proper motion and low dispersion measure result in such significant interstellar scintillation that detections with a high signal-to-noise ratio have required long observing durations or fortuitous timing. In this work, we turn to the sensitive MeerKAT telescope, and 
combined with historic Parkes data, are able to leverage the kinematic and relativistic effects of PSR J1933$-$6211 to constrain its 3D orbital geometry and the component masses. We obtain a precise proper motion magnitude of 12.42(3) mas yr$^{-1}$ and a parallax of 1.0(3)~mas, and we also measure their effects as secular changes in the Keplerian parameters of the orbit: a variation in the orbital period of $7(1) \times 10^{-13}$~s~s$^{-1}$ and a change in the projected semi-major axis of $1.60(5) \times 10^{-14}$~s~s$^{-1}$. A self-consistent analysis of all kinematic and relativistic effects yields a distance to the pulsar of $1.6^{+0.2}_{-0.3}$~kpc, an orbital inclination, $i = 55(1) \deg,$ and a longitude of the ascending node,  $\Omega = 255^{+8}_{-14}$ deg. The probability densities for $\Omega$ and $i$ and their symmetric counterparts, $180-i$ and $360-\Omega,$ are seen to depend on the chosen fiducial orbit used to measure the time of passage of periastron ($T_0$). We investigate this unexpected dependence and rule out software-related causes using simulations. Nevertheless, we constrain the masses of the pulsar and its companion to be $1.4^{+0.3}_{-0.2}$\,M$_\odot$ and $0.43(5)$\,M$_\odot$ , respectively. These results strongly disfavour a helium-dominated composition for the white dwarf companion. The similarity in the spin, orbital parameters, and companion masses of PSRs J1933$-$6211 and J1614$-$2230 suggests that these systems underwent case A Roche-lobe overflow, an extended evolutionary process that occurs while the companion star is still on the main sequence. However, \psr has not accreted significant matter: its mass is still at $\sim 1.4$ M$_\odot$.  This highlights the low accretion efficiency of the spin-up process and suggests that observed neutron star masses are mostly a result of supernova physics, with minimum influence of subsequent binary evolution.}

   \maketitle


\section{Introduction} \label{section:Introduction}

\psr was discovered as part of the Parkes High Galactic Latitude Survey \citep{Jacoby2007}, which used the 64 m CSIRO Parkes Murriyang radio telescope in Parkes, New South Wales, Australia (henceforth the Parkes telescope), to search for radio pulsars at Galactic latitudes between 15 and 30 $\deg$. The fully recycled nature of the pulsar, combined with a very low eccentricity ($e \sim 1.2 \times 10^{-6}$), indicates that the companion very likely is a white dwarf (WD) star, whose progenitor recycled the pulsar. \citet{Jacoby2007} used timing observations of the pulsar with the Parkes telescope and the CPSR2 backend to derive the binary mass function, and estimated a minimum mass for the companion ($\Mc$) of $0.32~\msun$ by assuming that the pulsar mass ($\Mp$) is $1.4~\msun$.

The fast spin of \psr ($P_0 = 3.5$\,ms) is typical of millisecond pulsars (MSPs). Most MSPs formed in low-mass X-ray binaries (LMXB), which allow for the long accretion times required to spin up neutron stars (NSs) to these short spin periods. In these systems, the companions are helium WDs (He WDs), in which the binary orbital period ($P_b$) and the WD mass are thought to be correlated \citep{TaurisAndSavonije1999}. For the orbital period of \psr (12.8 d), the correlation predicts a companion mass between 0.25 M$_{\odot}$ and 0.28 M$_{\odot}$, depending on the properties of the progenitor of the WD. 
This predicted value, using the He WD correlation, is lower than the originally estimated minimum companion mass ($0.32~\msun$), suggesting either an unusually light pulsar or that the companion is not a He WD, but a more massive type of WD, such as a carbon-oxygen (CO) WD, formed instead in an intermediate-mass X-ray binary (IMXB; \citealt{Tauris:2011ck}). 

These features make this system a relative rarity; there are only 4 other pulsars with spin rates below 6\,ms with established or likely CO-WD companions (PSRs J1101$-$6424; \citealt{Ng2015}; J1614$-$2230; \citealt{Alam2020}; J1618$-$4624; \citealt{Cameron2020}; and J1943+2210; \citealt{Scholz2015}), compared to the 101 pulsar binary systems in this spin period range with He WD companions \citep{ManchesterEtAl2005}.\footnote{See catalogue at: \href{http://www.atnf.csiro.au/research/pulsar/psrcat}{http://www.atnf.csiro.au/research/pulsar/psrcat}, version 1.67}
Because the more massive companions evolve faster, MSPs resulting from IMXBs tend to be significantly slower ($P\, > \, 9\, \rm ms$) than those that result from LMXBs. Thus, the spin periods of these pulsars are somewhat anomalous if the companion is a CO WD.

The orbital inclination of one of these systems, PSR~J1614$-$2230, is close to $90 \deg$, which allowed for a precise measurement of the Shapiro delay \citep{1964PhRvL..13..789S} and showed the pulsar to have a mass that is very likely above $1.9\, \msun$ \citep{DemorestEtAl2010,2018ApJS..235...37A}, thereby introducing strong constraints on the equation of state of dense nuclear matter \citep{Ozel&Freire2016}. This raises the question of how the pulsar becomes this massive. A detailed study of this system suggested an alternative evolutionary pathway, in which the NS was spun up via case A Roche-lobe overflow \citep[RLO; see][]{Tauris:2011ck}. Case A RLO takes place when the companion is still a main-sequence star. Under these conditions, the accretion timescale is related to the hydrogen-burning timescale of the donor.  This long accretion episode can therefore in principle allow the NS to gain much angular momentum, explaining the fast rotations observed in these systems. In contrast, in terms of mass gain, \cite{Tauris:2011ck} estimated that PSR~J1614$-$2230 gained  $\mathcal{O}(0.2)$\,M$_\odot$ at most during accretion, that is, they concluded that PSR~J1614$-$2230 is massive because it was born that way.

Motivated by the possibility that a CO WD companion might result in \psr being as massive as PSR~J1614$-$2230 (or by the possibility that if the companion is a He WD, the pulsar would have an unusually low mass), \cite{Graikou2017} attempted to measure the masses of the components of the \psr system via the Shapiro delay. They timed the system using coherently dedispersed Parkes data, which significantly improved the timing precision. A timing residual root-mean-square (rms) of 1.23 $\upmu$s was obtained using only a subset of bright observations in which the pulsar signal was boosted by interstellar scintillation. However, owing to unfavourable scintillation during many observations and the now-estimated low orbital inclination of the system, obtaining mass measurements was not possible. Based on their available data, the authors placed an upper limit of 0.44~$\msun$ on $\Mc$. 

In this work, we again attempt to measure the masses of the components of the \psr system; this time successfully. The superior sensitivity of the MeerKAT telescope \citep{Jonas2009} was crucial for this. It yields a sensitivity for pulsars that is an order of magnitude better than that of the Parkes telescope (\citealt{BailesEtAl2020}; the exact number depends on the spectral index of the pulsar and interstellar scintillation). This is especially important for pulsars like \psrc which are located so far south that they cannot be observed with any of the sensitive Northern Hemisphere telescopes. MeerKAT timing observations are carried out under the MeerTime Large Survey Project (LSP), pursuing a broad range of scientific topics \citep{BailesEtAl2020}. The \psr observations were made under two distinct research sub-themes within the MeerTime LSP: 1) the relativistic binary timing programme \cite[][RelBin]{Kramer2021}, which performs dedicated observations of binary pulsar systems to measure relativistic effects in their timing, with the aim of testing gravity theories and measuring NS masses, and 2) the pulsar timing array programme (\citealt{Spiewak2022}; PTA), which observes an array of southern millisecond pulsars to search for nanohertz gravitational waves. This paper reports the results of these timing measurements. Our results are aided by the aforementioned data from the Parkes telescope, and include additional measurements made with that telescope using the new Ultra-Wideband Low receiver \citep{HobbsEtAl2020}.

The structure of this paper is as follows. In section~\ref{sec:observations} we discuss the observations, the resulting data, and how these were analysed. In section~\ref{sec:prof} we present the polarimetric profile of the pulsar, together with a rotating vector model (RVM) of the polarimetry. In section~\ref{sec:timing} we present our timing results, in particular, a discussion of the most important timing parameters, the component masses, and the orbital orientation for this system. In section~\ref{sec:discussion} we discuss the main results and their implications for the nature of the system and for the evolution of MSP-CO WD systems in general. We also provide conclusions and future prospects.

\section{Observations and data reduction}
\label{sec:observations}

\begin{table*}[h]
 \caption[]{
 \label{tab:observing_details}
Observing systems and the timing data sets of \psr.}
 \centering
 \begin{threeparttable}
  \begin{tabular}[\textwidth]{p{0.5in} l l l l l l l l l}
   \hline
   \hline
   Telescope & Receiver & Backend & CF & BW & nchan & CD& Time span & Hours & \#ToAs\\
       &  &  &(MHz) & (MHz) &  & & (MJD) & observed\\
   \hline
   \multirow{4}{*}{Parkes} & 20-cm&CPSR2 & 1341/1405 & 2$\times$64 & $2\times128$ & No & 52795-53301 &11.2 & 70/64\\
   &multibeam&CASPSR & 1382 & 256 & 512 & Yes & 55676-56011 & 22.0 & 264\\
   &&CASPSR & 1382 & 256 & 512 & Yes &59139-59140 & 0.77 & 3\\
   & Ultra-Wide-  & Medusa& 2368 & 3328 & 3328 & Yes & 58336-59657 & 14.2 & 99\\
   & band Low &&&&& \\
   MeerKAT & L-band/1K & PTUSE & 1283.58 & 775.75 & 928 & Yes & 58550-59716& 24.5& 1016\\
   \hline
   Total & & & & & & & 6921 days & 72.7& 1516\\
   \hline
  \end{tabular}
 \tablefoot{Following the telescope, receiver, and backend specifications, we provide the associated centre observing frequency (CF), the effective observable bandwidth (BW), the number of frequency channels (nchan), whether intra-channel coherent dedispersion (CD) was applied, the time span, the hours observed, and the number of ToAs.}
 \end{threeparttable}
\end{table*}

In this work, we used the 2003/2004 time-of-arrival values (ToAs) from the Parkes observations described by \cite{Jacoby2007}, as well as 2011/2012 Parkes ToAs associated with the data in \citet{Graikou2017} and provided to us as 5 min time-averaged ToA values. These represent a curated ToA set from which outliers with a low signal-to-noise ratio (S/N) were removed. 
These ToAs were originally obtained using the Parkes 20-cm multibeam receiver \citep{Staveley-SmithEtAl1996} with the Caltech Swinburne Parkes Recorder 2 (CPSR2) and the CASPER Parkes Swinburne Recorder  \cite[][CASPSR]{VenkatramanKrishnanPhDThesis} backends. We describe the new observations we obtained below. Combined, the full timing baseline reported in this work is 19 years.
An overview of all the data and their characteristics used in this work is presented in Table~\ref{tab:observing_details}.

\subsection{Parkes observations}

More recent Parkes data of \psr were collected through the P965 and P1032 Parkes observing programmes. This includes coherently dedispersed fold-mode observations using the ultra-wide bandwidth low-frequency (UWL) receiver with its Medusa backend \citep{HobbsEtAl2020}; as well as a few fold-mode observations using the 20-cm multibeam receiver and the CASPSR backend. The latter setup is identical to the one used in the 2011/2012 data of \citet{Graikou2017} and therefore provides an overlap between the MeerKAT/PTUSE and Parkes/CASPSR data sets, which are otherwise separated by a large gap in observations from 2016 to 2019 that could hamper accurate phase connection for PSR J1933$-$6211. 

We have a total of 17 UWL observations that vary in duration from 890 sec to 1hr 4min, taken between August 6, 2018 and March 18, 2022. The Parkes UWL receiver operates at a centre frequency of 2368\, MHz and has a total bandwidth of 3328\,MHz.  In the fold-mode setup used here, it produces 1024 phase bins across the rotational phase of the pulsar.

\subsection{MeerKAT observations}

Data from the MeerKAT telescope were obtained between March 8, 2019 and May 16, 2022. The observations made by the PTA programme were 256\,s each and were regularly spaced, with a mean cadence of two weeks, while the RelBin observations were longer ($\geq$\,2048 seconds) and were aimed at obtaining good orbital coverage. In particular, the RelBin data set contains one 4\,hr observation (MJD 58746.80) and two 90\,min observations (MJDs 58836.50 and 58823.69) taken close to and across superior conjunction to optimise for Shapiro delay measurements.

The MeerKAT observations were recorded using the L-band receiver (856 - 1712 \,MHz) in its 1K (1024) channelisation mode, using the Pulsar Timing User Supplied Equipment (PTUSE) backend \citep{BailesEtAl2020}, which provided coherently dedispersed folded pulsar archives with 1024 phase bins across the pulse profile of 3.54\,ms, or with a phase-bin resolution of 3.46 $\upmu$s.

Prior to the observations, standard array calibration is applied via the MeerKAT science data processing (SDP) pipeline, as described in \citet{Serylak2021}. This includes online polarisation calibration, such that (since April 9, 2020) the Tied Array Beam data stream ingested to PTUSE produces polarisation-calibrated L-band pulsar data products. Data recorded before access to the online polarisation calibration pipeline were calibrated offline according to the steps outlined in \citet{Serylak2021}.

\subsection{Data reduction}

The data reduction and analysis in this section rely on a combination of well-established pulsar software suites, including \textsc{psrchive} \citep{HotanEtAl2004} and \textsc{tempo2} \citep{HobbsEtAl2006,EdwardsEtAl2006}, as well as observatory or research programme-specific pipelines (e.g. \textsc{meerpipe}, \textsc{psrpype}). We denote particular tools within software packages as \texttt{tool}/\textsc{software}.

\subsubsection{Parkes: Multibeam/CASPSR}\label{sec:dataredparkes}

The CASPSR data taken in October 2020 were reduced in a similar manner as reported in \citet{Graikou2017}. Band edges were removed and radio frequency interference (RFI) manually excised using \texttt{pazi}/\textsc{psrchive} before creating frequency integrated, intensity-only (Stokes I) profiles with 512 phase bins using \texttt{pam}. 

Testament to the scintillating nature of \psrc of the four observations obtained (with observing lengths ranging from $\sim 1 - 1.5$\,hr), only the two observations taken on 17 October produced profiles with S/N>10. The brightest of these were reduced to two time intervals, and the second to a single averaged profile only. 

\subsubsection{Parkes: UWL/Medusa}\label{sec:datareduwl}

Data from the UWL receiver were reduced using the \textsc{psrpype} processing pipeline \footnote{\url{https://github.com/vivekvenkris/psrpype}}. The pipeline performs flux and polarimetric calibration, along with automated RFI excision using \textsc{clfd}\footnote{\url{https://github.com/v-morello/clfd}}. This works in a similar way to \textsc{meerpipe} and produces RFI excised, calibrated, and decimated to a number of time, frequency, and polarisation resolutions. To increase the profile S/N values leading up to computed ToA measurements, we further reduced the data products to four frequency channels, single time integrations, and full intensity only. 

\subsubsection{MeerKAT: L-band/1K PTUSE}\label{sec:dataredMKT}

The MeerTime observations were reduced using the \textsc{Meerpipe} pipeline\footnote{\url{https://bitbucket.org/meertime/meerpipe/src/master/}}, which produces archive files cleaned from RFI (based on a modified version of \textsc{coastguard;} \citealt{LazarusEtAl2016}) of varying decimation using standard \texttt{pam}/\textsc{psrchive} commands. We started our customised data reduction from the output products containing 16 frequency channels across the inner 775.75 MHz of MeerKAT L-band,  an eight-fold reduction in subintegration time, and calibrated Stokes information. 

Based on our findings that the \psr scintillation cycles last approximately 20 to 30 minutes on average, all longer-duration observations were decimated to have a minimum integration length of 500 seconds. To increase the S/N per ToA, we reduced the channelisation to eight frequency channels for all data. A rotation measure (RM) correction of 9.2 rad m$^{-2}$ was applied using \texttt{pam}, based on the measurement presented in \citet{Kramer2021}.

\subsection{Estimating pulse times of arrival}\label{sec:ToAcreation}

Three additional CASPSR ToAs were added to the data set using the same standard profile as was used to generate the CASPSR ToAs in \citet{Graikou2017}, together with the data described in Sect. \ref{sec:dataredparkes}. This provided an overlap between the Parkes/CASPSR and MeerKAT/PTUSE ToAs.  

To create ToA values from the UWL/Medusa data sets, a high S/N timing standard was created through the addition of the available observations (\texttt{psradd}/\textsc{psrchive}) following their reduction and cleaning by \textsc{psrpype} as well as additional RFI removal by hand using \texttt{pazi}.  Based on the obtained S/N values, we chose to create a total intensity standard with four frequency channels (providing a per channel profile with an S/N 70 to 350) and turned them into DM-corrected analytical templates using \texttt{psrsmooth}/\textsc{psrchive}. This template was then used in \texttt{pat}/\textsc{psrchive} to obtain ToA values at the telescope for the reduced UWL data described in Sect. \ref{sec:datareduwl}.

A MeerKAT multi-frequency timing standard was created using all \psr observations with an estimated S/N$>200$. These were added using \texttt{psradd} and reduced to create a template with a single subintegration, eight frequency channels with four Stokes polarisations, with DM and RM corrections applied.  Polarisation-resolved standards were motivated by the timing improvements \citet{Graikou2017} reported using matrix template matching (MTM; \citealt{vanStraten2004,vanStraten2013}) for ToA generation.

Analytical polarisation and frequency-resolved standards were generated from these high S/N templates by again applying wavelet smoothing using \texttt{psrsmooth}. These were subsequently used to apply MTM using \texttt{pat} on the MeerKAT data products described in Sect. \ref{sec:dataredMKT}, providing measurements of the ToAs. As shown by \cite{Graikou2017}, the timing of \psr benefits especially from using the MTM because of its sharp polarisation features, as shown in Fig. \ref{fig:profpa}. Use of MTM strongly relies on an accurate polarisation calibration of the pulsar data. The well-calibrated MeerTime data products will therefore benefit from the use of MTM. 

Finally, to account for the varying S/N values of the observations that are due to the high occurrence of scintillation in \psrc we manually removed individual ToAs with uncertainties larger than 20 $\upmu$s from all data sets. This was done after visual inspections that confirmed that the large ToA uncertainties were indeed due to low S/N detections.

\subsection{Timing analysis and orbital models}\label{sec:timingmodel}

The analysis of the ToAs was made using \textsc{tempo2}\footnote{\href{https://bitbucket.org/psrsoft/tempo2/src/master/}{https://bitbucket.org/psrsoft/tempo2/src/master/}}.
The telescope-specific ToAs computed above were transformed into TT(BIPM2021)\footnote{https://webtai.bipm.org/ftp/pub/tai/ttbipm/TTBIPM.2021}, which is a realisation of terrestrial time as defined by the International Astronomical Union (IAU), and thereafter converted into time of arrivals at the Solar System barycentre using the most recent DE440 Solar System ephemeris of the Jet Propulsion Laboratory (JPL; \citealt{Park2021}).

Initial orbital and pulsar parameter estimates were found using the DDH orbital model description as implemented by the \textsc{tempo2} software. This is an extension of the DD model \citep{DD2}, and describes the Keplerian orbit via the parameters orbital period ($P_{\rm b}$), length of the projected semi-major axis ($x_{\rm p}$), orbital eccentricity ($e$), longitude of the periastron ($\omega$), and the time of passage through the ascending node ($T_0$)  along with several relativistic corrections, which are quantified by a set of phenomenological post-Keplerian (PK) parameters. In particular, DDH uses the orthometric amplitude ($h_3$) and the orthometric ratio ($\varsigma$) to model the Shapiro delay, whereas the standard DD model describes it with the range ($r$) and shape ($s$) parameters \citep{FreireAndWex2010, Weisberg&Huang2016}. These parameters have the advantage of being far less strongly correlated than $r$ and $s$, especially for low orbital inclinations, as is the case for \psrf

However, the DDH model fails to account for the full set of kinematic contributions described in Sect. \ref{sec:kin}; in particular, it does not describe the annual orbital parallax (AOP; \citealt{Kopeikin1995}), but can only model the secular variation of $x$ caused by the proper motion \citep{1996ApJ...467L..93K}, $\dot{x}$. Consequently, it cannot discriminate between the multiple solutions for the orbital orientation of the system given by a measured $\dot{x}$ and $\varsigma$. Furthermore, unmodeled residual trends caused by the AOP pollute the very weak Shapiro delay signal whose higher harmonics are of the same order of magnitude as the AOP for \psrf 

For these reasons, we refined our parameter estimations by using the T2 binary model, which is based on the DD model, but self-consistently accounts for all kinematic contributions to orbital and post-Keplerian parameters described in Sect. \ref{sec:kin} \citep{EdwardsEtAl2006}. Within the description of the T2 model, all kinematic effects caused by the proper motion are calculated internally from the orbital orientation of the system, given by the position angle of the ascending node ($\Omega$, KOM) and orbital inclination ($i$, KIN) parameters. If astrometric dynamics is the only cause of the variation of the semi-major axis, then there is no need for an additional $\dot{x}$ parameter under this paradigm.

We note that for systems with very low orbital eccentricities, such as \psrc $\omega$ and $T_0$ estimated through the DD or T2 model, for example, can be highly correlated. The ELL1-type orbital models \citep{Lange2001} are a popular alternative to replace these with the time of ascending node ($T_{\rm asc} = T_0 -\omega\, P_b$) and the Laplace-Lagrange parameters, $\epsilon_1 \equiv e \sin \omega$ and $\epsilon_2 \equiv e \cos \omega$. Similarly to the DD models, however, the ELL1-type orbital models fail to include the relevant kinematic contributions included in the T2 model.  Consequently, following our T2 analysis, we derived $T_{\rm asc}, \epsilon_1$ and $\epsilon_2$ to produce a full set of accurate timing parameters. We note that the \textsc{tempo2} implementation of the T2 model can also work with the ELL1 parametrisation, which we also performed as a check and obtained consistent results.

In order to calculate reliable error bars and parameter correlations within the T2 model, we employed the \textsc{temponest} plugin to \textsc{tempo2}. \textsc{Temponest} is a Bayesian parameter estimation tool that allows for physically motivated prior distributions on timing parameter values while also fitting for additional noise models to the data, including red noise and DM noise \citep{LentatiEtAl2014}. \textsc{Temponest} internally uses the \textsc{MultiNest} \citep{Feroz2019} sampler. We set the multi-modal flag ON, as we expected multiple modes to be present for some of our parameters a priori.

\subsection{Noise model selection}
\label{sec:noise_models}

We settled on a best noise model to describe the \psr timing data by performing Bayesian non-linear fits of timing models with varying noise characteristics to the data using \textsc{temponest}. The tested noise models included 1) a white-noise only model, where we fit for the noise parameters EFAC and EQUAD that add to or scale the uncertainties of the ToA measurements (as described in \citealt{LentatiEtAl2014}); 2) white noise plus a DM noise model characterised through a chromatic power-law model and; 3) white noise plus a stochastic achromatic (red) timing-noise model similarly described by a power law; as well as 4) white noise plus DM noise plus a red-noise model. To each model, we provided uniform priors centred on the initial best-fit tempo2 parameter value and ranging across $\pm40\sigma$, where $\sigma$ is the associated tempo2 uncertainty (i.e. we set FitSig to 40 in \textsc{temponest}). For a select set of parameters, we provided physically motivated uniform priors; ($\Omega$, KOM) and ($i$, KIN) were set to cover their range of possible values: [0,360] deg and [0,180] deg., respectively; 
($\varpi$, PX) was set to range from 0.1 to 2.2 and ($M_c$, M2) from to 0.1 to 1.5~M$_{\odot}$.

We performed Bayes factor (BF) comparisons between these models, and we find the strongest evidence for a red- and white-noise model, which compared to white-noise only has a BF of 16.6. Comparisons of the red- and white-noise model to models that include DM-noise yield a BF of 5.4 against DM and white noise and 1.8 against DM, white, and red noise.
We conclude that all DM effects are well modelled through the inclusion of the \textsc{tempo2} timing parameters, DM1 and DM2 (which describe the coefficients to the first- and second-order DM derivatives expressed as a DM Taylor series expansion). For the remainder of the results section, we therefore focus on the outcomes of the \textsc{temponest} posterior distributions, which include red- and white-noise parameters. The amplitude and power-law spectral index of the red noise is provided in Table~\ref{tab:timing2}.

\section{Results: Profile analysis}\label{sec:prof}

\begin{figure*}
    \centering
    \includegraphics[width=0.75\textwidth, trim={25 25 25 25}]{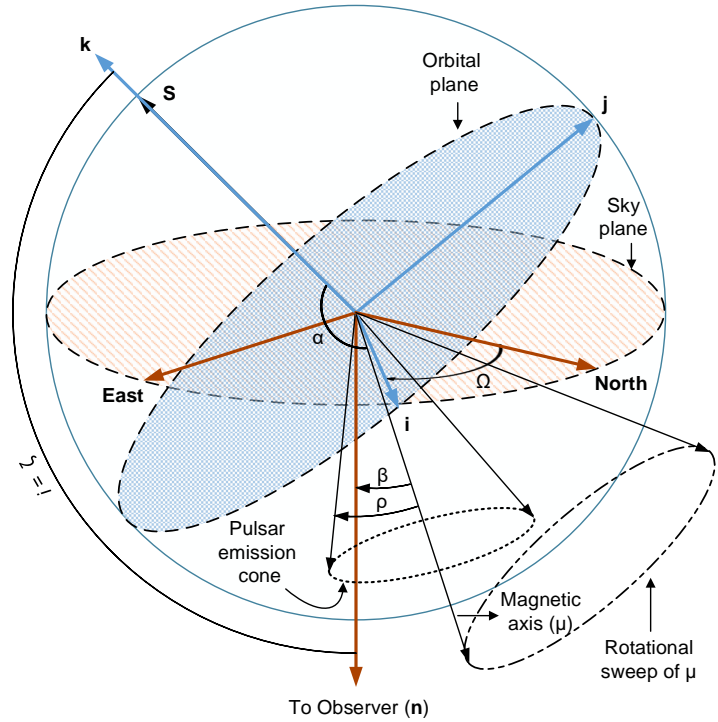}
    \caption{Schematic of the 3D geometry for the \psr binary in the observer's convention. The orbital plane is defined by the unit vectors {\bf i} and {\bf j}, with the orbital angular momentum perpendicular to the plane, along the direction of {\bf k}. The spin of the pulsar {\bf S} is assumed to be aligned with the orbital angular momentum. The plane of the sky is defined by the {\bf east} and {\bf north} unit vectors, with {\bf n} being the unit vector of our line of sight that is defined from the pulsar to the observer. The angle between {\bf n} and {\bf k}  by definition is the orbital inclination angle, $i$. In the observer's convention, this is equal to the polar angle of the pulsar $\zeta$. The projection of {\bf S} on the sky plane forms. The angle between {\bf north} and $\bf i$ is the longitude of the ascending node. $\rho$ denotes the opening angle of the emission cone, with $\beta$ the impact angle of $n$ on the cone, as measured from the magnetic axis $\mu$. $\alpha_{\rm p}$ is the inclination of $\mu$ with respect to {\bf S}.  }
    \label{fig:3d}
\end{figure*}

Throughout this paper, we use the observer's convention to define angles and vectors, unless explicitly stated otherwise. In this framework, the position angle and the longitude of the ascending node ($\Omega$) increase counter-clockwise on the plane of the sky, starting from north. Furthermore, the orbital inclination $i$ is defined as the angle between the orbital angular momentum and the line from the pulsar to the Earth. This can vary between 0 and $180\, \deg$. Fig. \ref{fig:3d} shows these angular definitions. The observer's convention is also used by the T2 orbital model \citep{EdwardsEtAl2006}. Any angle without a subscript follows this convention. We recall that this convention is different from the conventions used in \citet{DamourTaylor1992} and \citet{Kopeikin1995}, where $\Omega$ was measured clockwise from east and $i$ is the angle between the angular momentum of the orbit and a vector pointing from the Earth to the pulsar. Angles in this alternate convention are explicitly denoted with the subscript DT92.

\begin{figure}
    \includegraphics[width=0.95\columnwidth]{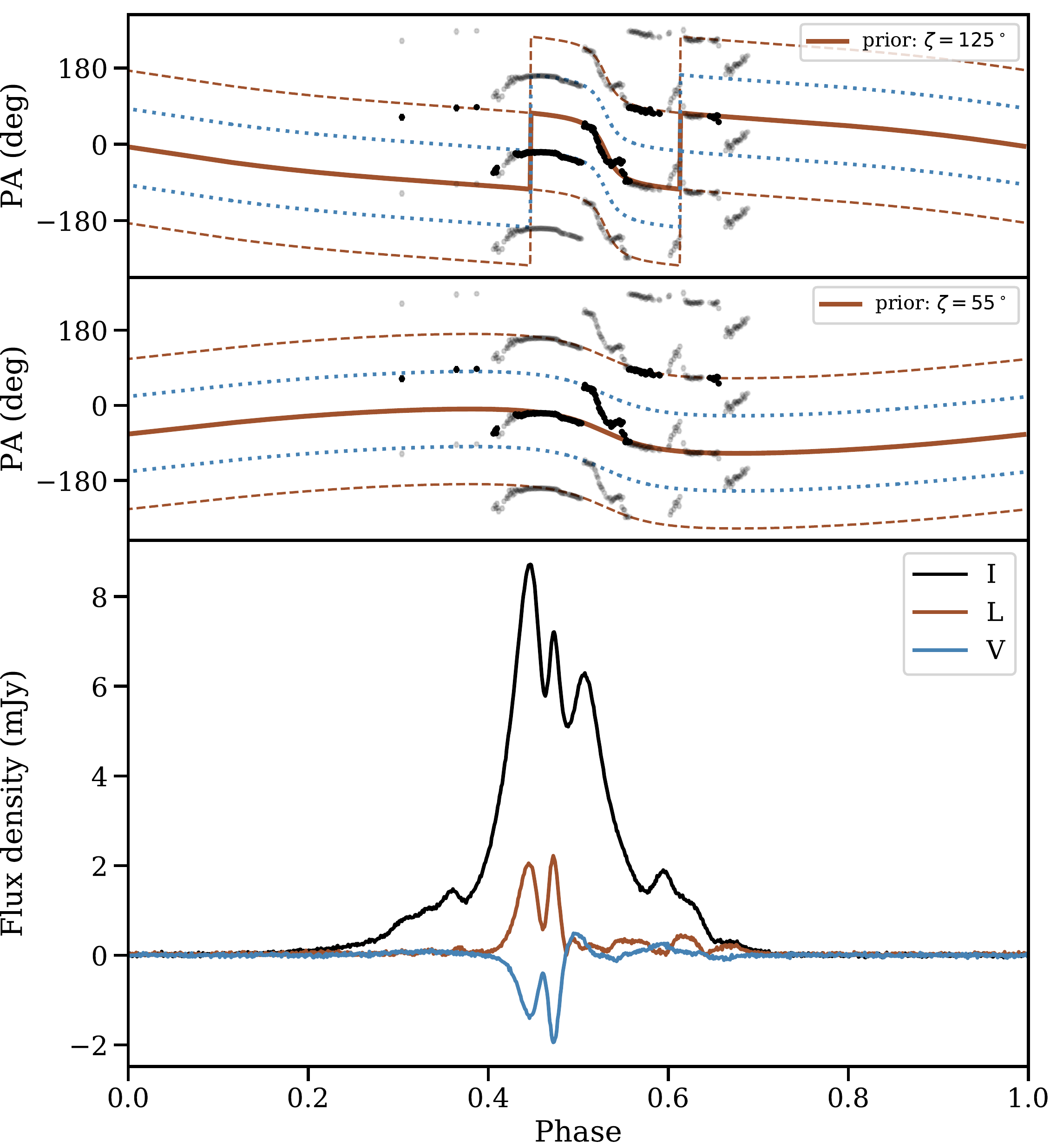}
    \caption{Averaged flux-calibrated and RM-corrected \psr pulse profile (\textit{bottom}) and associated polarisation position angles (\textit{middle and top}). \textit{Bottom:} This averaged flux-calibrated and RM-corrected profile was obtained from adding all 24.5 hours of MeerKAT L-band data, providing an \mbox{$\rm{S/N} \approx 3000$}, a peak flux density of 8.7~mJy, and a mean flux density of 1.1~mJy. The profile shows significant linear (red, L) and circular (blue, V) polarisation, the sharped-edged features of which make it an ideal candidate for MTM timing techniques when detecting it brightly. \textit{Middle and top:} The linear polarisation PA are plotted for all phase bins with significant (> 5$\sigma$) linear polarisation. To visualise PA swings, we present the PA values, together with PA+180 deg and PA-180 deg (grey data points) in both panels. The solid (brown) line shows the best-fit RVM model when using a prior of $\zeta = 125$ deg and obtaining $\alpha = 41.66(4)$ deg  (\textit{top}), or a prior of $\zeta = 55$ deg and obtaining $\alpha = 121.99(4)$ deg (\textit{middle}) as described in Sect.~\ref{sec:rvm}, with their $\pm 180$ deg equivalent as dashed (brown) lines. This fit was obtained using only the data points in black and removing the fainter data points of the centrally plotted PA values. The dotted lines (blue) in both panels are PA$\pm 90$ deg to show the potential PA values of orthogonally polarised modes.}
    \label{fig:profpa}
\end{figure}

Fig.~\ref{fig:profpa} provides our highest S/N profile for \psr as obtained when adding 88079\,sec (24.5 hr) MeerKAT L-band data, cleaned from RFI. The flux-calibrated and RM-corrected profile has a mean flux density of 1.1~mJy and an estimated \mbox{$\rm{S/N} = 2833$}.

\subsection{Pulsar geometry using pulse structure data}\label{sec:rvm}

The variation in position angle of the linear polarisation (PA; $\psi$) of the pulse profile  across the pulsar
longitude changes due to the viewing geometry, and under ideal assumptions, it results in an S-shaped swing.  
This is often described by the rotating vector model \citep[][RVM]{Radhakrishnan&Cooke}, which can then provide information about the pulsar geometry. The RVM describes $\psi$ as a function of the pulse phase, $\phi$, depending on the magnetic inclination angle, $\alpha,$ and the viewing angle, $\zeta$, which is the angle between the line-of-sight vector and the pulsar spin and can be written as
\begin{equation}
\label{eqn:rvm}
{\rm \psi} = {\rm \psi}_{0} +
{\rm arctan} \left( \frac{{\rm sin}\alpha
\, {\rm sin}(\phi_0 - \phi)}{{\rm sin}\zeta
\, {\rm cos}\alpha - {\rm cos}\zeta
\, {\rm sin}\alpha \, {\rm cos}(\phi_0 - \phi)} \right),
\end{equation}

where we have modified the equation to follow the observer's convention.

Many studies have shown that deviations from the RVM model are typical especially for MSPs (e.g. \citealt{Yan2011}, \citealt{Dai_2015}), and we therefore do not expect good agreement with the RVM model for PSR J1933$-$6211. However, in particular cases, such as for MSP PSR J1811$-$2405, the PA values follow an RVM model, which has proven effective in breaking the $x_p \sin i$ degeneracy in Shapiro delay measurements to obtain an accurate orbital inclination \citep{Kramer2021}.

Our obtained PA values for \psrc shown in the top and middle panels of Fig.~\ref{fig:profpa}, clearly exhibit more complex variations than the simple RVM S-shaped swing described above. The sharp change in slope of the PA points, especially towards the rear end of the profile, suggests that the orthogonally polarised modes are mixed,  which makes some PA points unreliable.

Within our plotted PA values, S-shaped curves are discernible, and we therefore 
attempt to fit Eq.~\eqref{eqn:rvm} to select PA values after removing points that deviated from an RVM-like swing. A blind fit of the remaining points after accounting for a PA jump from orthogonally polarised modes (at phase 0.43) and with a flat prior on $\alpha$ and $\zeta$, provides a surprisingly precise value for $\zeta$ of 34(1)~deg. The posterior distribution of $\alpha$ is bimodal at both 36(1)~deg and 34(1)~deg. These values are consistent with the similar analysis of \cite{Kramer2021}, according to the DT92 convention. 

For systems in which the spin of the pulsar is expected to be aligned with the orbital angular momentum (e.g. PSR J1933$-$ 6211), $\zeta \equiv i \equiv 180-i_{\rm DT92}$. However, our timing measurement of the inclination angle (see Sect.~\ref{sec:timing}) is inconsistent with $\zeta$ (or 180-$\zeta$). This confirms that the PA swing indeed follows more complex variations than can be explained by RVM. 

As an additional check, we set $\zeta \equiv i$ and $\zeta \equiv 180-i$ (where $i$ was obtained from timing) as the prior and performed constrained fits. These provided an $\alpha$ value of 41.66(4) and 121.99(4) deg, respectively.  The corresponding RVM curves are shown in the middle and top panels of Fig.~\ref{fig:profpa}. However, a Bayes factor test between the blind and constrained fit shows that the blind fit is strongly favoured (BFs > 400). We conclude that even for the curated PA points that seem to follow an S-type curve, their variations do not follow the RVM.

\section{Results: Timing analysis} \label{sec:timing}

\begin{table*}
\caption{\psr timing parameters obtained from \textsc{temponest}.
}
\centering 
\begin{tabular} {l l}
\hline
\hline
\multicolumn{2}{c}{Observation and data reduction parameters}\\
\hline
Timing model\dotfill & T2\\
Solar System ephemeris\dotfill & DE440\\[0.5ex]
Timescale \dotfill &  TT(BP2021)\\[0.5ex]
Reference epoch for period, position and DM (MJD)\dotfill & 58831\\[0.5ex]
Solar wind electron number density, $n_{0}$ (cm$^{-3}$)\dotfill & 9.961\\[0.5ex]
\hline
\multicolumn{2}{c}{Spin and astrometric parameters}\\
\hline
Right ascension, $\alpha$ (J2000, h:m:s)\dotfill &  19:33:32.413992(9)\\
Declination, $\delta$ (J2000, d:m:s)\dotfill & $-$62:11:46.70233(9)\\[0.5ex]
Proper motion in $\alpha$, $\mu_{\alpha}$ (mas\,yr$^{-1}$)\dotfill &  $-$5.62(1)\\[0.5ex]
Proper motion in $\delta$, $\mu_{\delta}$ (mas\,yr$^{-1}$)\dotfill  & 11.09(3)\\[0.5ex]
Parallax, $\varpi$ (mas)\dotfill&  1.0(3)\\
Spin frequency, $\nu$ (Hz)\dotfill & 282.212313459989(3)\\[0.5ex]
Spin-down rate, $\dot{\nu}$ ($10^{-16}$\,Hz\,s$^{-1}$)\dotfill & $-$3.0830(2)\\[0.5ex]
Dispersion measure, DM (cm$^{-3}$\,pc)\dotfill  &  11.507(3)\\[0.5ex]
First Derivative of DM, DM1 (cm$^{-3}$\,pc\,yr$^{-1}$)\dotfill  & 0.00032(3)\\[0.5ex]
Second Derivative of DM, DM2 (\,cm$^{-3}$\,pc\,yr$^{-2}$)\dotfill  & $-$0.00033(1)\\[0.5ex]
Rotation measure, RM (rad\,m$^{-2}$)\dotfill  & 9.2(1)$^{(a)}$\\[0.5ex]
\hline
\multicolumn{2}{c}{Derived parameters}\\
\hline
Galactic longitude, $l$ ($^{\circ}$)\dotfill  &334.4309\\
Galactic latitude, $b$ ($^{\circ}$)\dotfill  &-28.6315\\[0.5ex]
Total proper motion, $\mu_{\text{T}}$ (mas\,yr$^{-1}$)\dotfill & 12.42(3)\\[0.5ex]
DM-derived distance (NE2001), $d$ (kpc)\dotfill &  0.51\\[0.5ex]
DM-derived distance (YMW16), $d$ (kpc)\dotfill &  0.65\\[0.5ex]
Parallax derived distance, $d$ (kpc)\dotfill &  1.0(3)\\[0.5ex]
Parallax derived distance including EDSD prior, $d$ (kpc)\dotfill &  1.2$^{+0.5}_{-0.3}$\\[0.5ex]
$\dot{P_{\rm{b}}}$-derived distance, $d$ (kpc)\dotfill &  1.7(3)\\[0.5ex]
Parallax distance including EDSD prior, $d$ (kpc)\dotfill &  1.4(2)\\[0.5ex]
Distance derived from combining parallax, \pbdot and EDSD prior, $d$ (kpc)\dotfill & 1.6$^{+0.2}_{-0.3}$\\[0.5ex]
Spin period, $P_{0}$ (ms)\dotfill &  3.5434314957408(4)\\[0.5ex]
Spin period derivative, $\dot{P}$ ($10^{-21}$\,s\,s$^{-1}$)\dotfill & 3.8710(2)\\[0.5ex]
Total kinematic contribution to $\dot{P}$, $\dot{P}_{\text{k}}$ ($10^{-21}$\,s\,s$^{-1}$)\dotfill & -1.6(3)\\[0.5ex] 
Intrinsic spin period derivative, $\dot{P}_{\text{int}}$ ($10^{-21}$\,s\,s$^{-1}$)\dotfill & 2.2(3)\\[0.5ex] 
Inferred surface magnetic field, $B_{\text{surf}}$ ($10^{7}$\,G)\dotfill & $\sim$ 9.3\\[0.5ex]
Inferred characteristic age, $\tau_{\text{c}}$ (Gyr)\dotfill & $\sim$ 24\\[0.5ex]
Inferred spin-down luminosity, $\dot{E}$ ($10^{33}$\,erg\,s$^{-1}$)\dotfill &$\sim$2.11\\[0.5ex]
\hline
\hline

\label{tab:timing}
\end{tabular}
\tablefoot{
Results were obtained using the \textsc{tempo2} T2 binary model within \textsc{temponest}. We also present derived computed quantities from these. All uncertainties in the measured values are 68.3\,\% confidence limits (C.~L). Binary and post-Keplerian parameters are presented in Table \ref{tab:timing2}.\\
\tablefoottext{a}{As obtained in \citet{Kramer2021}.}}
\end{table*}

\begin{table*}
\caption{Binary timing parameters and associated mass and inclination values for PSR~J1933$-$6211.}
\centering 
\begin{tabular} {l l}
\hline
\hline
\multicolumn{2}{c}{Keplerian parameters}\\
\hline
Orbital period, $P_{\text{b}}$ (days)\dotfill & 12.819406716(1)\\[0.5ex]
Projected semi-major axis of the pulsar orbit, $x_p$ (s)\dotfill & 12.2815670(5)\\[0.5ex]
Epoch of periastron, $T_{0}$ (MJD)\dotfill & 53004.13(2)\\[0.5ex]
Orbital eccentricity, $e$ (10$^{-6}$)\dotfill & 1.26(2)\\[0.5ex]
Longitude of periastron at $T_{0}$, $\omega$ ($^{\circ}$)\dotfill &102.1(5)\\[0.5ex]

\hline
\multicolumn{2}{c}{Post-Keplerian parameters and orbital geometry}\\
\hline
Orbital period derivative, $\dot{P_{\text{b}}}$ ($10^{-13}$\,s\,s$^{-1}$)\dotfill & 7(1)\\[0.5ex]
Rate of change of orbital semi-major axis, $\dot{x}$ ($10^{-14}$\,s\,s$^{-1}$)\dotfill &$1.60(5)^{(a)}$\\ [0.5ex]
Range of Shapiro delay, $r$ ($T_\odot$)\dotfill & $0.43(5)$\\[0.5ex]
Longitude of the ascending node , $\Omega$ (deg)$^\dagger$\dotfill &$255^{+8}_{-14}$\\[0.5ex]
Orbital inclination, $i$ ($\deg$)$^\dagger$\dotfill & 55(1)\\[0.5ex]
\hline
\multicolumn{2}{c}{Noise parameters}\\
\hline
EFAC MeerKAT L-band/1K \dotfill & 0.80\\
EFAC Parkes CASPSR\dotfill  & 0.85\\
EFAC Parkes CPSR2 1341 MHz\dotfill  & 0.65\\
EFAC Parkes CPSR2 1405 MHz\dotfill  & 0.80\\
EFAC Parkes UWL \dotfill & 1.2\\
Log$_{10}$[EQUAD(s)] MeerKAT L-band/1K \dotfill & $-$6.4\\
Log$_{10}$[EQUAD(s)] Parkes CASPSR\dotfill  & $-$8.2\\
Log$_{10}$[EQUAD(s)] Parkes CPSR2 1341 MHz\dotfill  & $-$8.5\\
Log$_{10}$[EQUAD(s)] Parkes CPSR2 1405 MHz\dotfill  & $-$8.1\\
Log$_{10}$[EQUAD(s)] Parkes UWL \dotfill & $-$6.1\\
Red noise power-law amplitude, $A_{red}$ \dotfill  & $-13.01(8)$\\ [0.5ex]
Red noise power-law spectral index, $\alpha_{red}$\dotfill & $1.8(4)$\\ [0.5ex]
\hline
\multicolumn{2}{c}{Mass and inclination measurements}\\
\hline
Mass function, $f$ (M$_{\odot}$)\dotfill & 0.0121034266(2) \\[0.5ex]
Companion mass, $M_{\text{c}}$ (M$_{\odot}$)\dotfill & $0.43(5)$\\[0.5ex]
Pulsar mass, $M_{\text{p}}$ (M$_{\odot}$)\dotfill & $1.4^{+0.3}_{-0.2}$ \\[0.5ex]
\hline
\multicolumn{2}{c}{Derived parameters}\\
\hline
Orthometric amplitude, $h3$ (10$^{-7}$)\dotfill & $3.0(4)$\\[0.5ex]
Orthometric ratio, $\stig$ \dotfill & 0.52(1)\\[0.5ex]
Time of an ascending node passage, $T_{\rm ASC}$ (MJD)\dotfill & 53013.31465961(7)\\[0.5ex]
Laplace-Lagrange parameter, $\epsilon_1 =e\sin\omega$ (10$^{-6}$) \dotfill & 1.22(2)\\[0.5ex]
Laplace-Lagrange parameter, $\epsilon_2 = e\cos\omega$ (10$^{-6}$)\dotfill &$-0.272(8)$\\[0.5ex]
Contribution to $\dot{P_{\text{b}}}$ from Shklovskii effect$^\ddagger$, $\dot{P_{\text{b}^{\text{Shk}}}}$ ($10^{-13}$\,s\,s$^{-1}$)\dotfill & $5^{+2}_{-1}$ \\
Contribution to $\dot{P_{\text{b}}}$ from Galactic rotation$^\ddagger$, $\dot{P_{\text{b}^{\text{gal}}}}$ ($10^{-13}$\,s\,s$^{-1}$)\dotfill & $-0.3^{+0.2}_{-0.1}$ \\
\hline
\hline

\label{tab:timing2}
\end{tabular}
\tablefoot{
The results are obtained from our \textsc{temponest} outputs of the \textsc{tempo2} T2 binary model. Derived orthometric Shapiro delay and ELL1 timing parameters are also computed for reference. All uncertainties in the measured values are 68.3\,\% confidence limits.
\tablefoottext{a}{As obtained using the DDH binary model.}
\tablefoottext{b}{$\Omega$ and $i$, should be read within the context of Sect.~\ref{sec:T0dependence}.}
\tablefoottext{c}{Obtained with the EDSD distance prior, see Sect.~\ref{sec:dist}.}

}
\end{table*}

\begin{figure*}
    \includegraphics[width=\textwidth]{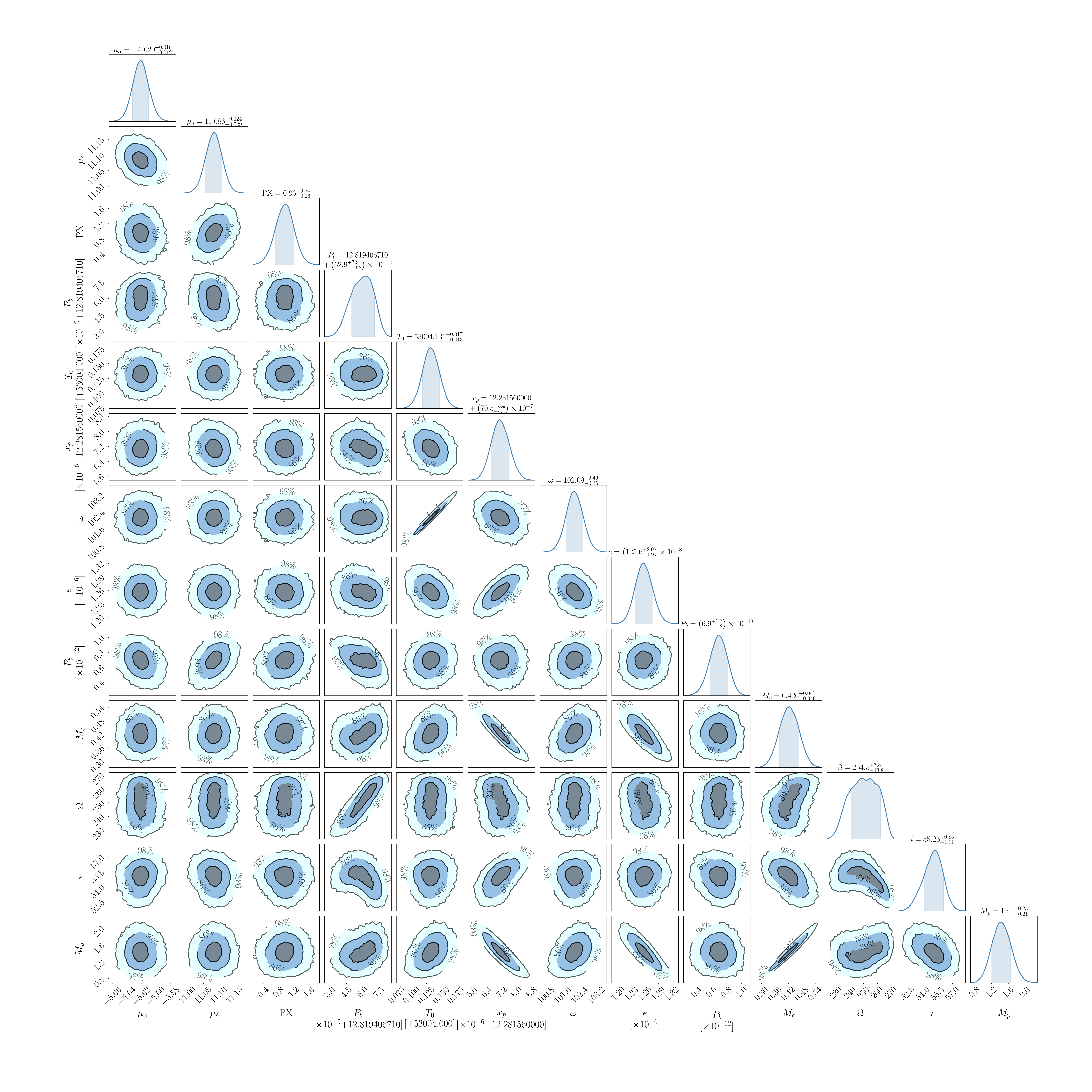}
    \caption{Output posterior distributions for \psr parameters.  These were obtained from \textsc{temponest} sampling using the T2 orbital model and a red-noise model, and were generated using the \textsc{chainconsumer} package. We present the orbital and post-Keplerian parameters. Full details are provided in Table \ref{tab:timing}. The obtained pulsar mass ($M_p$) distribution was computed using the mass function and the posterior distributions on $M_c$, $i$, $x_p$ , and $P_b$.}
    \label{fig:corner}
\end{figure*}

\begin{figure*}
    \centering
    \includegraphics[width=\textwidth]{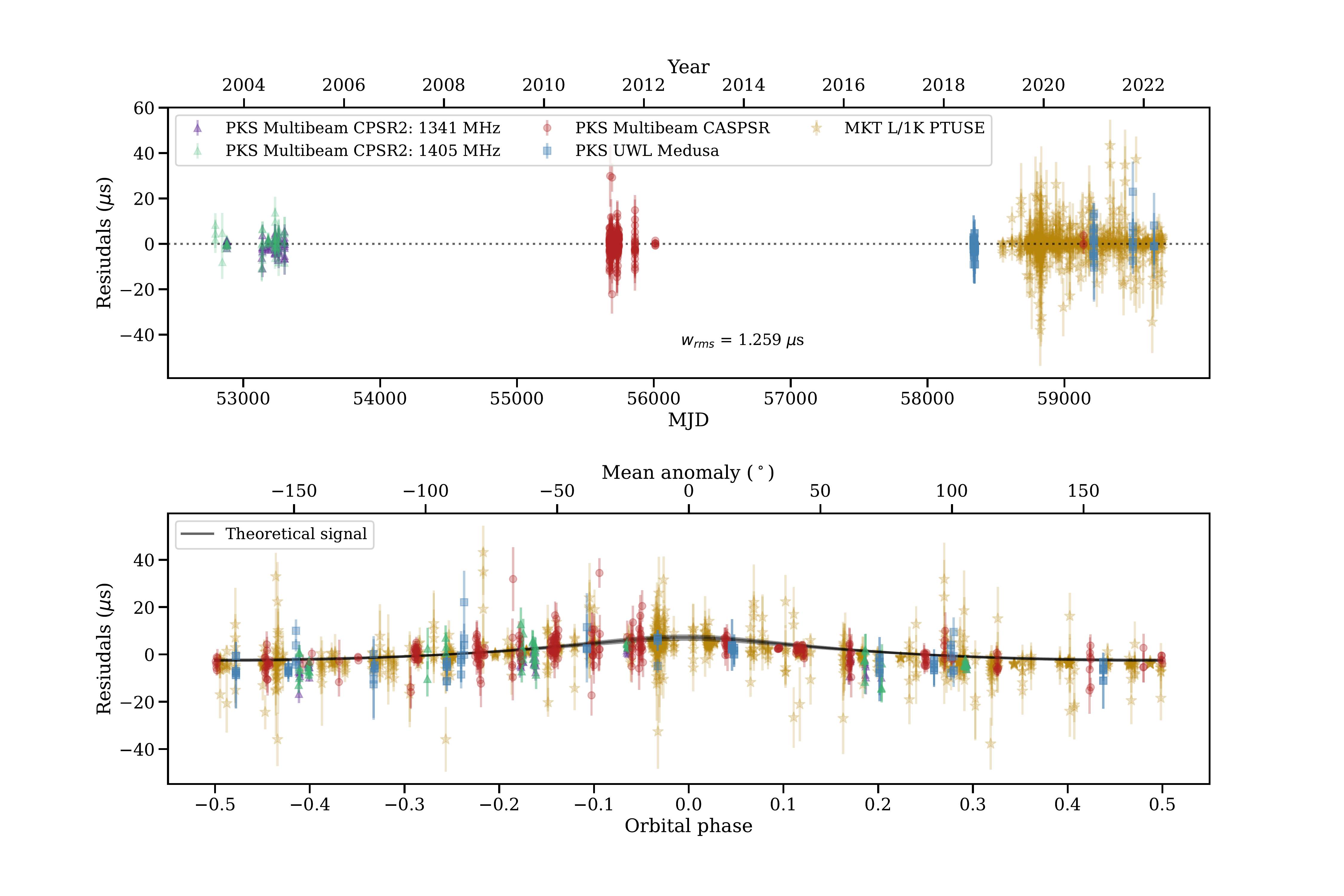}
    \caption{Timing residuals across epochs (\textit{top}) and orbital phase (\textit{bottom}). Data include historic Parkes ToAs (green and purple; PKS Multibeam CPSR2 1341\,MHz and 1405\,MHz) recent and historic Parkes CASPSR data (red; PKS Multibeam CASPSR), as well as recent Parkes UWL/Medusa ToAs (blue; PKS UWL Medusa) and MeerKAT L-band ToAs (gold; MKT L/1K PTUSE). \textit{Top:} After applying the best-fit values as obtained using the timing and noise model described in Sect. \ref{sec:timingmodel}, we obtain a weighted rms of 1.259 $\upmu$s. \textit{Bottom:} Residuals as a function of orbital phase, measured from superior conjunction (the ascending node is at a phase of $-0.25$). 
    A Shapiro delay signal is discernible when setting the companion mass to zero ($M_c = 0$) while keeping all other parameters fixed, including $x_p \sin i$ . We overplot the expected theoretical signal based on the best-fit inclination and companion mass values in black. The line width indicates combined 1$\sigma$ deviations in companion mass and inclination angle.}
    \label{fig:res}
\end{figure*}

The complete set of spin and astrometric timing parameters using the T2 model is provided in Table \ref{tab:timing}, while the measured binary parameters are contained in Table \ref{tab:timing2}. We also present a set of derived quantities in both tables, which include the values for timing parameters used in ELL1-type orbital models ($T_{\rm asc}, \epsilon_1$ and $\epsilon_2$). These are useful to record fold-mode data for the pulsar. 

We used the \textsc{chainconsumer} library \citep{Hinton2016_ChainConsumer} to visualise the \textsc{temponest} T2 posterior distributions, with corner plots showing the 1D and 2D posterior distributions of the parameters. Fig. \ref{fig:corner} shows the resulting output for a subset of timing parameters of interest. Here, we obtained the red-noise model in \textsc{temponest} with 5000 live points to produce well-sampled distributions. Parameter error bars are 1$\sigma$ uncertainties following the default smoothing as applied through \textsc{chainconsumer}. 

The solution presented in Tables~\ref{tab:timing} and \ref{tab:timing2} provides a good description of the timing data. In the top panel of Fig. \ref{fig:res}, we show the timing residuals having implemented this best-fit model as a function of the observing date and observing system (see the figure caption for a description of the colouring). The timing residuals show the difference between the observed barycentric ToA value (obtained using the techniques described in Sect. \ref{sec:ToAcreation}) and the predicted barycentric arrival time for that particular pulsar rotation based on the single best-fit timing model as obtained above. The validity of the timing model is evident from the low weighted rms ($w_{\rm rms} = 1.372\upmu$s) of the residuals, the obtained reduced $\chi^2$ value of 0.99. There appears to be no unmodelled trends in the residuals.

In subsequent sections, we highlight a few of the physically interesting parameter results obtained from the timing analyses and resulting posterior distribution, in particular, some of the astrometric parameters (including parallax and distance estimates) and the PK parameters that allow for estimates of the component masses and orbital orientation of the system.

\subsection{Proper motion}

Our updated position and proper motion values for \psr provide an improvement in precision by a factor of 8 compared to the values published in \citet{Graikou2017}.
From the measured proper motion values in right ascension and declination ($\mu_\alpha, \mu_\delta$), we obtain a total proper motion magnitude value of 12.42(3) mas yr$^{-1}$. The corresponding position angle of the proper motion, \mbox{$\Theta_\mu = \tan^{-1}\,(\mu_\alpha/\mu_\delta)$}  is $333.1(1)$ deg in the observer's convention (see Sect.~\ref{sec:prof} and Fig.~\ref{fig:3d}).

\begin{figure*}
\centering
    \includegraphics[width=0.75\textwidth]{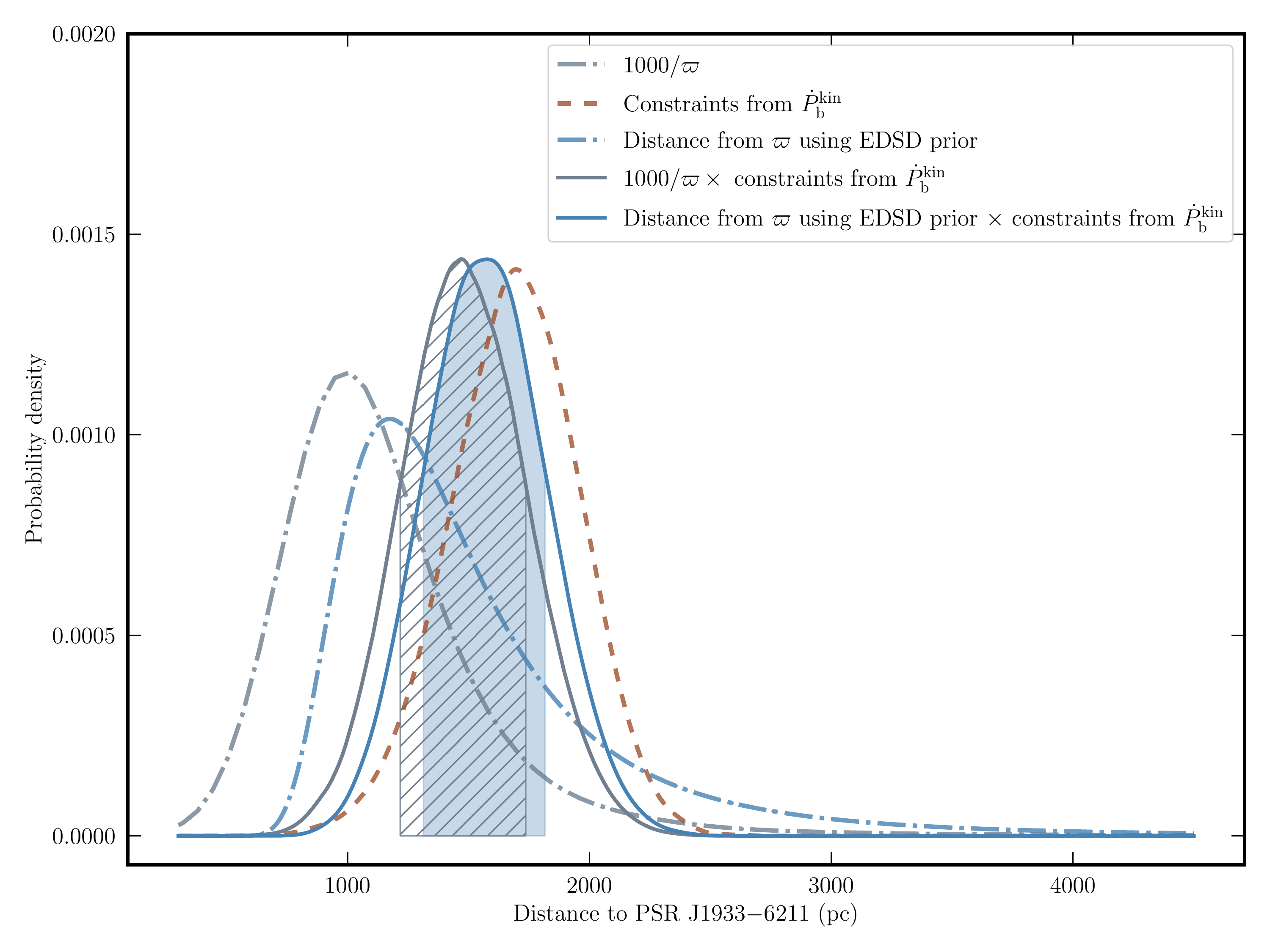}
    \caption{Constraints on the distance to PSR J1933$-$6211. The dash-dotted grey and blue lines indicate the constraints from the measurement of timing parallax without and with  
    corrections for Lutz-Kelker bias using an EDSD prior, respectively. The dashed brown line indicates constraints from the kinematic contribution to the measured orbital period decay ($\dot{P}_{\rm b}^{\rm kin}$). The solid grey line shows the combined constraints from a simple inversion of the timing parallax and $\dot{P}_{\rm b}^{\rm kin}$, with grey hatched lines marking the 68\% C.~L. This corresponds to a distance of 
    $1.4(2)$ kpc. The solid blue line shows the combined constraints from the timing parallax with the EDSD prior and $\dot{P}_{\rm b}^{\rm kin}$, with the solid blue shaded region denoting its 68\% C.~L. This corresponds to a distance of $1.6^{+0.2}_{-0.3}$ kpc.}
    \label{fig:kin}
\end{figure*}

\subsection{Kinematic effects on the pulsar timing parameters}\label{sec:kin}

The moderate distance of \psr and the combination of relatively large proper motion, large projected semi-major axis of its orbit ($x_{\rm p}$), relatively low orbital inclination of $i \sim 55\, \deg$ and high timing precision provide a rare combination of criteria that enable the detection of subtle kinematic effects that help constrain the 3D geometry of the system. These effects, first described in detail by \citet{Kopeikin1995,1996ApJ...467L..93K}, must be modelled precisely; otherwise, the unaccounted-for delays will pollute our measurement of the weak Shapiro delay in this system. This is a consequence of the small $i$.

We now describe these effects in more detail. They depend on the absolute orientation of the system, which is given by the position angle of the line of nodes (the intersection of the orbital plane with the plane of the sky), $\Omega$, and the orbital inclination, $i$.

\subsubsection{Proper motion contributions to \omdot and \xdot}

The high proper motion of the \psr binary leads to a constant change in the viewing angle of the pulsar, which manifestd as a constantly changing longitude of periastron ($\omega$) and orbital inclination $i$; the latter might measurably change $x$, which is given by $x_{\rm p} \sin i /c$, even if the semi-major axis of the pulsar orbit ($x_{\rm p}$) does not actually change. In the observer's convention, these kinematic contributions to $\dot{x}$ and $\dot{\omega}$ are given by 

\begin{align}
    \dot{x}_\mu & =  1.54\times10^{-16} \,\, x \,\, \cot i \,\, (-\mu_\alpha \cos \Omega + \mu_\delta \sin \Omega) \,\,\rm{s\,s}^{-1}\label{eq:xdotmu}\\
    \dot{\omega_\mu} & = 2.78\times10^{-7} \,\, \csc i \,\, (\mu_\alpha \cos \Omega + \mu_\delta \sin \Omega)\,\,\rm{deg/yr}. \label{eq:omegadotmu}
\end{align}

\noindent The expression in Eq.~\eqref{eq:xdotmu} is identically to Eq.~(1) in \citet{Guo2021}, for example, which provides a convention-independent alternative. We note that for Eq.~\eqref{eq:xdotmu} to become valid in the DT92 convention, the angles need to be transformed accordingly, with $\Omega = 90^{\circ} - \Omega_{DT92}$ and $i = 180^{\circ} - i_{DT92}$.

Given the low orbital eccentricity, we do not measure a significant $\dot{\omega}$. However, we measure a highly
significant $\dot{x}$, $1.60(5) \times 10^{-14}$~s~s$^{-1}$ assuming the DD model. A detailed analysis of all possible contributions to $\dot{x}$ (e.g. \citealt{Lorimer&Kramer2005}) shows that this must be almost exclusively caused by the proper motion according to Eq.~\eqref{eq:xdotmu}. For this reason, the measured $\dot{x}$ leads to constraints on the orbital orientation of the system, that is, $\Omega$ and $i$ (see Sect.~\ref{sec:shapiro} for details).

\subsubsection{Annual orbital parallax}

The variation in the Earth's position as it orbits the Sun causes small annual changes to $x_{\rm p}$ (from the apparent change in the orbital inclination caused by the Earth's motion) and $\omega$. This effect, termed the AOP, is generally very small. However, it is the key for determining the absolute orbital orientation of the system. The reason is that if we measure $\sin i$ from Shapiro delay and $\dot{x}$ for a particular binary, Eq.~\eqref{eq:xdotmu} yields four possible solutions for $\Omega$ and $i$. Measuring the impact of the annular orbital parallax through $\dot{x}$ allows us to ultimately break this degeneracy. 

This cyclic effect of AOP, which has variations at both orbital and annual timescales, imprints itself on $x$ and $\omega$ and can be expressed as in \citet{Kopeikin1995},
\begin{equation}
 \Delta_\pi = -\frac{1}{c d}\, \Big(\vec{r}\,\cdot \vec{r_p} - (\mathbf{K_0}\,\vec{r})\,\, \,(\mathbf{K_0}\,\vec{r_p})\,\Big), \label{eq:Kop95}
\end{equation}

\noindent with $c$ the speed of light, and $d$ the distance between the binary and the SSB. The vectors $\vec{r} = (X,Y,Z)$ and $\vec{r_p}$ describe the Earth's position with respect to the SSB and the pulsar position with respect to the SSB, respectively. The unit normal vector $\mathbf{K_0}$, points from the SSB to the barycentre of the binary. The values of $\vec{r}$ and $\vec{r_p}$ will depend on the Solar System ephemeris model that is employed, and they vary with time.  

Following the expressions in \citet{Kopeikin1995}, we simplify Eq.~\eqref{eq:Kop95} to obtain an estimate on the expected peak-to-peak amplitude of the AOP. In doing so, we make the simplifying assumption that both the pulsar's binary orbit and Earth's orbit are circular ($e = 0$) and find   

\begin{align}\label{eq:aop}
\Delta_\pi  = \frac{x_p}{d}\,\Big[&( \Delta_{\mathbf{I_0}} \sin \Omega - \Delta_{\mathbf{J_0}} \cos \Omega)\,\, \sin (\omega_{\rm PB}\, t) \cot{i} \\ - \,
&( \Delta_{\mathbf{I_0}} \cos \Omega + \Delta_{\mathbf{J_0}} \sin \Omega )\,\, \cos \nonumber (\omega_{\rm PB}\,t) \csc{i} \Big], 
\end{align}

\noindent where $x_{\rm p}$, $\Omega,$ and $i$ are as before (and given in Table~\ref{tab:timing2}), and $\omega_{\rm PB} = 2\pi/P_b$ is the binary orbital frequency. 

The unit vectors ($\textbf{I}_0$,\, $\textbf{J}_0$, \,$\textbf{K}_0$), describe the coordinate system of the pulsar reference frame, with its origin at the binary system barycentre. Following \citet{Kopeikin1995}, 
\begin{align}
     \Delta_{\textbf{I}_0} &= (\vec{r} \cdot \textbf{I}_0) = -X \sin\alpha + Y \cos \alpha \\
     \Delta_{\textbf{J}_0} &= (\vec{r} \cdot \textbf{J}_0) = -X \sin \delta \cos \alpha - Y\sin \delta \sin \alpha, 
\end{align}

\noindent with ($\alpha$,$\delta$) the right ascension and declination of \psrc and $\vec{r} = (X, Y, Z)$ as before. Using the same current JPL solar ephemeris as in our timing results (DE440), which is contained within the \texttt{jplephem} package and implemented in \texttt{astropy}, we obtain the Earth's $(X, Y, Z)$ coordinates as a function of our observing MJD range. 

We next use Eq.~\eqref{eq:aop} to compute the resulting $\Delta_\pi$ oscillatory trend as a function of MJD and find a peak-to-peak orbital parallax of \psr of $\sim100$ ns. 

Table \ref{tab:timing} shows that the precision of $x_p$, following our timing timing analysis, is of the order of 50 ns, so that an AOP contribution per ToA ranging from approximately -50 to 50 ns will have a measurable and time-dependent impact on $x$. The importance of using the T2 model to account for this AOP and its contribution to \xdot is re-emphasised by this comparison. 

\subsubsection{Distance estimates from \pbdot and $\varpi$ }
\label{sec:dist}

We measure a decay of the orbital period of $7(1) \times 10^{-13}$s\,s$^{-1}$, as presented in Table~\ref{tab:timing2}. This \pbdot  measurement can arise from a number of contributing effects, 
\begin{equation}
    \dot {P_b} =\dot {P_b}^{\rm GR} + \dot {P_b}^{\rm kin} + \dot {P_b}^{\dot{m}} +\dot {P_b}^{\rm T}  
,\end{equation}

where the terms indicate contributions due to gravitational wave decay (GR), kinematic contributions due to changing Doppler shift (kin), mass loss in the system ($\dot{m}$), and tidal dissipation of the orbit. We find that the only non-negligible contribution for \psr arises from the kinematic contributions, which consist of two secular acceleration effects,

\begin{equation}
   \left( \frac{\dot {P_{\rm b}}}{P_{\rm b}} \right) ^{\rm kin} = \left( \frac{\dot{P}_{\rm b}}{{P}_{\rm b}} \right) ^{\rm Shk}  + \,\, \left( \frac{\dot{P}_{\rm b}}{{P}_{\rm b}} \right)^{\rm Gal}.
    \label{eq:pbdot_kin}
\end{equation}

Here, $\dot {P_b}^{\rm shk}$ is the acceleration due to transverse motion, also known as the Shklovskii effect, and $\dot {P_b}^{\rm Gal}$ is the acceleration of the binary in the gravitational field of the Milky Way due to differential rotation. $\dot {P_b}^{\rm shk}$ depends on the transverse proper motion of the pulsar ($\mu_{\rm T}$) and the distance to the pulsar ($d_{\rm PSR}$) and is related by

\begin{equation}
    \left( \frac{\dot{P}_{\rm b}}{P_{\rm b}} \right)^{\rm Shk}   = 2.43 \times 10^{-21} \left( \frac{\mu_{\mathrm T}}{\rm{mas \, yr^{-1}}}\right)^2 \left( \frac{d_{\rm PSR}}{\rm{kpc}} \right).
\end{equation}

 $\dot{P}_{\rm b}^{\rm Gal}$ also depends on $d_{\rm PSR}$, along with a rotation model for the Galaxy that provides the position of the Solar System and the pulsar with respect to the Galactic barycentre, and their relative accelerations. To compute the planar and azimuthal Galactic contribution to \pbdot , we use 
 
 \begin{align}
     \left( \frac{\dot{P}_{\rm b}}{P_{\rm b}} \right)^{\rm Gal\, pl}  &= 
     - \frac{1}{c} \frac{\Omega_\odot^2}{R_\odot} \left(\cos l + \frac{\beta}{\beta^2 + {\sin^2 l}}      \right)\, \cos b\\
     \left( \frac{\dot{P}_{\rm b}}{P_{\rm b}} \right)^{\rm Gal\, az} &=  - \frac{K_z  \mid {\sin b} \mid} {c}, 
\end{align}
 
as in \citet{LazaridisEtAl2009}, and implemented in the \textsc{GalDynPsr} library \citep{Pathak_2018}. Here, $(l, b)$ are the Galactic coordinates of the pulsar, and $\beta \equiv (d/R_\odot)\cos b - \cos l$, $R_\odot = 8.275$~kpc, and $\Omega_\odot = 240.5 \rm~km~s^{-1}$ are the Galactic distance of Earth and the orbital velocity. Current estimates of these parameters can be obtained from \citet{McMillan2017}, where $K_z/c$ is the vertical component of Galactic acceleration, 

\begin{equation}
    \frac{K_z}{c} = -1.08\times10^{-19} \, \left[0.58 + \frac{1.25}{(z_{\rm kpc}^2 + 0.0324)^{1/2}}  \right]\,z_{\rm kpc},
\end{equation}

with $z_{\rm kpc}$ the vertical height of the pulsar in kiloparsec (\cite{Pathak_2018}). 
 
Since both the Shklovskii and the Galactic acceleration effects depend linearly on $d_{\rm PSR}$, we can use the \pbdot measurement to provide a constraint on the pulsar distance independent of the more standard distance constraint obtained from the parallax measurements \citep{Bell&Bailes1996}.
 
From our timing analysis, we have a direct measurement of the pulsar parallax of $\varpi = 1.0(3)$. A simple inversion of this measurement provides a distance estimate of $1.0(3)$ kpc. However, given the low $\sim5\sigma$ significance of the measurement, this simple inversion is prone to the Lutz-Kelker bias \citep{LutzKelker1973} of exponentially increasing stellar density with distance. We corrected for this bias using a scaled probability density function following \cite{Antoniadis2021} \citep[see also][]{Verbiest:2012kh,Bailer-Jones:2018abc,Jennings:2018psk},

\begin{equation}
    P(d_{\rm PSR} | \varpi) = \frac{1}{2L^3}e^{-d_{\rm PSR}/L}d^2e^{(-1/d_{\rm PSR} -\varpi)^2/2\sigma_{\varpi}^2}.
\end{equation}

Here, we adopted an exponentially decreasing space density (EDSD) prior to avoid the divergence issues implicit in the original  Lutz-Kelker correction \citep[see][for details]{Bailer-Jones:2018abc}. $L$ can be thought of as a characteristic length scale,  which we set equal to $1.35$\,kpc, following \cite{Antoniadis2021}. 
The estimate of the distance corrected for the L-K bias from the timing parallax is $1.2^{+0.5}_{-0.3}$ kpc.
 
As described above, we can also obtain an additional distance estimate from the kinematically dominated \pbdot value. We used model C of the \textsc{GalDynPsr} library \citep{Pathak_2018} to evaluate Eqs. (10) to (13) together with the current values of $R\odot$ and $\Omega_\odot$ given above, to compute all kinematic contributions to \pbdot.  We obtain a distance estimate of $1.7^{+0.2}_{-0.3}$ kpc. 

This is consistent with the L-K corrected distance estimate from parallax. We also combined the probability densities of the distance estimates from $\varpi$ and \pbdot to obtain a more constraining distance of $1.4(2)$ kpc and $1.6^{+0.2}_{-0.3}$ kpc without and with correcting for the L-K bias, respectively. Fig. \ref{fig:kin} provides the PDF of the distance constraints for all these considerations. 

Comparing these distance estimates to the DM-based distance estimates of the NE2001 and YMW16 electron density models, which predict 510 pc and 650 pc, respectively, we find that both electron density models significantly underestimate the distance along this line of sight. We note that discrepancies between DM-estimated distances and parallax-inferred distances are common (e.g. \citealt{Stovall2019}), especially for high Galactic latitudes, and that independent distance measurements serve to improve electron density models for particular lines of sight.

\subsection{Shapiro delay, masses, and orbital orientation}
\label{sec:shapiro}

As Fig. \ref{fig:res} shows, the Shapiro delay signal in this pulsar has a maximum of only 7.16 $\upmu$s, a consequence of the far-from-edge-on configuration, and the reason why this delay was not detected until now. From the DDH model, we can estimate $\stig$ and $h_3$ from this signal. Combining this with the measurement of \xdot, we obtain constraints on the orbital inclination $i$ and $\Omega$. These are depicted graphically in Fig.~\ref{fig:cosi}, where the constraints from $\stig$ are presented by the dotted black lines and the constraints from \xdot are presented by the brown lines. According to these DDH obtained values, two possible solutions exist, one solution with $i \sim 55 \deg,$ and a second solution with $i \sim 125 \deg$.

Similarly, the T2 binary timing model can be used to obtain Shapiro delay estimates (see Table~\ref{tab:timing2}). However, this model simultaneously takes into account the effect of the AOP, the importance of which will become clear below.

We performed a full \textsc{temponest} analysis of the parameters in the T2 model. The resulting parameter uncertainties and their correlations are shown in Fig.~\ref{fig:corner}. 

The associated values for $\Omega$ and $\cos i$ are shown as blue shaded contours in Fig.~\ref{fig:cosi}. Comparing these with the constraints derived from the DDH model, we see that the degeneracy between the two possible $i$-$\Omega$ solutions is lifted: The solution at $i \sim 125\, \deg$ is excluded by the measurement of the AOP within the T2 model, which constrains $\Omega$ to \mbox{$255^{+8}_{-14} \, \deg$}. Within this narrow window, the inclination range is better constrained by \xdot than by $\stig$; a consequence of this is that the uncertainty on $\cos i$ in the top panel is significantly narrower than the 1$\sigma$ uncertainty of $\stig$, resulting in an unusually precise measurement of $i$, $55(1)\deg$. These estimates are for $T_0 \simeq 53004.16$. The reason we specify this will become evident below.

\subsubsection{Dependence of the orbital orientation estimates on T0}
\label{sec:T0dependence}

We observe an unexpected dependence of the constraints derived for the 3D orientation of the pulsar (i.e. $i$ and $\Omega$) on the fiducial orbit that we chose to measure $T_0$ (or equivalently, $T_{\rm asc}$). While we describe the changes only with respect to $T_0$ in the following, we observe a similar dependence using $T_{\rm asc}$, using the ELL1 formulation within the T2 model.

\begin{figure*}
    \centering
    \includegraphics[width=0.75\textwidth]{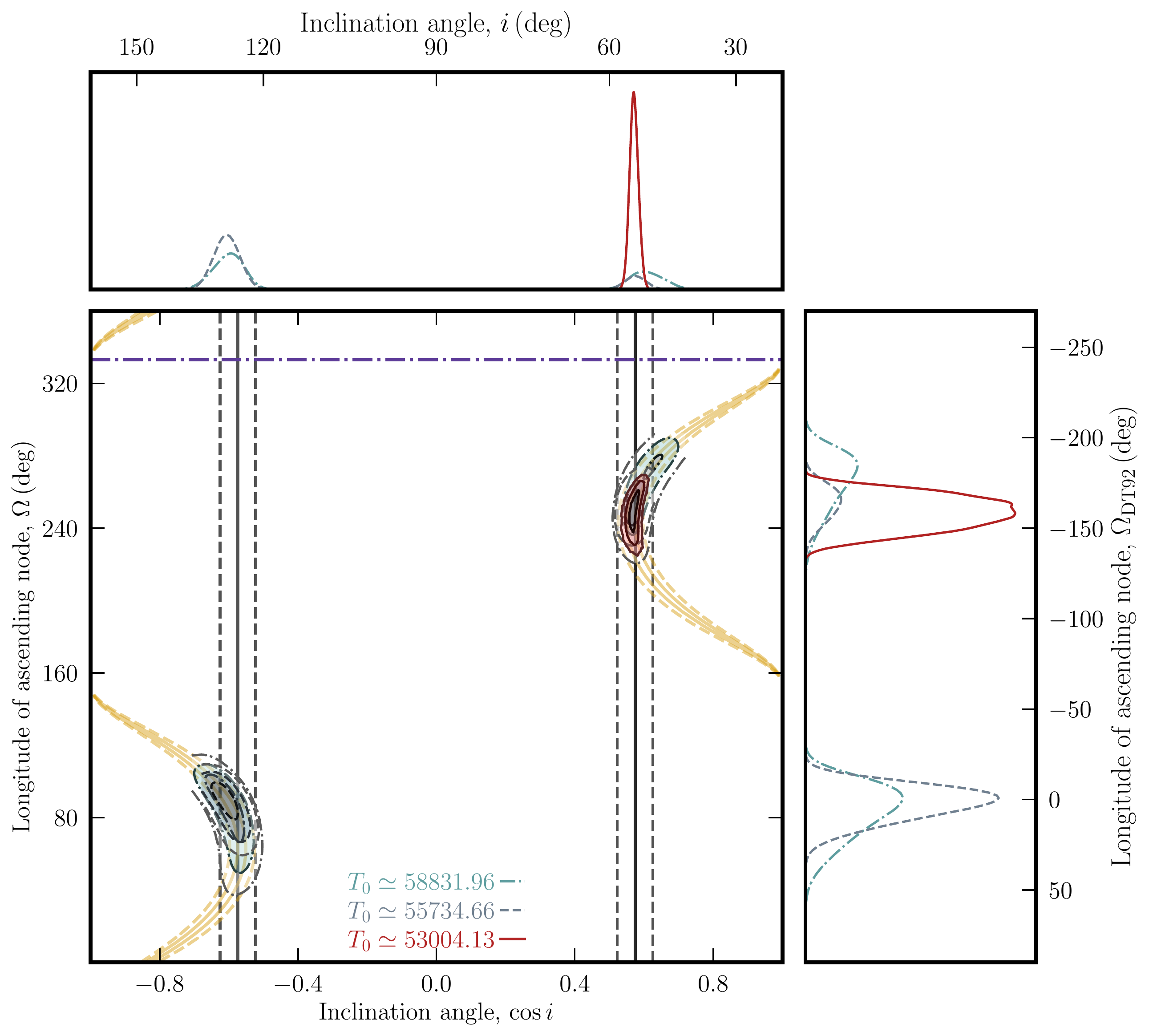}
    \caption{Allowable orbital orientation for \psr binary.  In the main panel, we display the orbital orientation space in the observer's convention (the $\Omega$ according to the DT92 convention is shown by the scale on the right, the $\cos i$ in that convention is simply $- \cos i$). The lines represent constraints from two parameters measured with a DDH solution: in black we show the constraint from $\varsigma$, and in brown we show the constraint from $\dot{x}$ (1$\sigma$ solid, 3$\sigma$ dashed) assuming it is solely caused by the proper motion of the system; the position angle of the latter is shown by the dot-dashed purple line. The contours, enclosing progressively darker shades of red, include 98\%, 86\% and 39\% C.~L. of the 2D probability density function (pdf) and represent the T2 timing model solutions calculated by \textsc{temponest} and as presented in Table \ref{tab:timing2}. This model takes into account all kinematic effects, including the AOP. The marginalised constraints on $\cos i$ and $\Omega$ are shown as 1D histograms (in red) in the  the top and side panels, respectively. For this model, with $T_0 \sim 53004$, the AOP is clearly detected, as we can see from the fact that the degeneracy between the two possible $\cos i$ - $\Omega$ solutions has been lifted: no probability remains in the solution at $\Omega \sim 60 \deg$, $\cos i \sim -0.6$. We also include the $\cos i$ - $\Omega$ probability density contours as obtained with $T_0$ set to 55734 (dashed grey) and 58836 (dot-dashed blue), respectively. As described in Sect.~\ref{sec:T0dependence}, we unexpectedly observe $i$ and $\Omega$ estimates to depend on $T_0$.} \label{fig:cosi}
\end{figure*}

What appears to be a significant detection of AOP at $T_0 \simeq 53004.13$, strong enough to entirely rule out the other $\Omega-\cos i$ island in Fig.~\ref{fig:cosi}, becomes less significant for $T_0$ values set to the later epochs of the data set. Of the three distinct observing campaigns on the pulsar (see Fig. \ref{fig:res}), the $T_0$ value in Table \ref{tab:timing2} is roughly in the middle of the first campaign with the CPSR2 backend, conducted soon after discovery. We repeated all the analyses with $T_0 \simeq 55734.66$ at the centre of the CASPSR data taken around 2011 and $T_0 \simeq 58836.96,$ which is the middle of our latest, largest, and most sensitive dataset from the MeerKAT L-band and the Parkes UWL receivers. The corresponding posteriors of $i$ and $\Omega$ are also shown in Fig. ~\ref{fig:cosi}, where the reduction in our sensitivity to AOP is evident. We rigorously tested whether these dependences were due to our software implementations by performing simulations that we detail in Appendix \ref{sec:app}. We also repeated our analysis of the data with twice the number of \textsc{temponest} live points (i.e. 10 000) for $T_0 \simeq 53004.13$ and $55734.66$ to understand whether we sufficiently sampled the global minima. While doing this, we extended the initial prior range for the parameters without physically motivated custom priors  from $\pm 40 \sigma$ to $\pm 100\sigma$ (the parameters with custom priors already had liberal prior distributions; see Sect. \ref{sec:noise_models}). This ensured that we sampled a larger parameter space and that our solutions were indeed the global minima. We find results consistent with Fig.~\ref{fig:cosi}, and for $T_0 \simeq 53004.13$ consistent with Table \ref{tab:timing2}. Based on these results and the simulations, we conclude that we do not find strong evidence that the  $T_0$ dependence is caused by the timing software or the analysis method.

This leaves the tantalising possibility that this is indeed physical, which we do not fully understand. The fact that regardless of $T_0$, we obtain probability islands in the same quadrants as in Fig.~\ref{fig:cosi} validates the robustness of our measurement of $\dot{x}$. All other parameters are seen to be almost identical across all the $T_0$ runs. The nominal proper motion of the system combined with a long orbital period negates the need for any additional or higher-order corrections to the astrometric and relativistic parameters other than what is already modelled by the T2 model. Hence, the physical origin of the dependence of AOP on $T_0$ is currently unclear. However, because our simulations suggest that we might be able to consistently obtain the 3D position for a similar data set with the same cadence and noise properties (see Appendix A for more details), we chose the $T_0$ value that provided the most constraints on the 3D geometry for Table \ref{tab:timing2}.

\subsubsection{Self-consistent mass measurements}

Regardless of the sense of $i$, its precise measurement
means that the weak Shapiro delay signal is used solely to determine the companion mass, $\Mc = 0.43(5) \, \msun$, that is, no precision is lost because of the correlation between $i$ and $\Mc$. From the mass function and the precise $i$ and $\Mc$, we find a pulsar mass of $M_p = 1.4^{+0.3}_{-0.2}$ M$_\odot$. This mass measurement is consistent for all $T_0$ values.

The orbital models used in these analyses are independent of theory; however, we know from many other experiments \citep{BertottiEtAL2003,2011MNRAS.412.2763F,Guo2021} that for weakly gravitating objects such as the Sun or WD stars, the Shapiro delay constraints from the Shapiro delay parameters can be translated directly into the constraints on $M_c$ and $i$. In addition, the constraints from \xdot and the AOP are purely geometric, such that our \textsc{temponest} analysis with the T2 model yields $M_c$, $i$, and $\Omega$ directly, without the need for further assumptions on the theory of gravity used, as would be required if additional PK parameters had been measured.

\subsection{Testing the time variation of the gravitational constant with \psr}

The fact that our measurement of \pbdot is consistent with almost entirely resulting from kinematic contributions together with an independent pulsar distance measured by the timing parallax allowed us to perform a test of the rate of change in the (local) gravitational constant ($\dot{G}$) over the time span of our observations. This change in $G$ is predicted by several classes of alternative theories of gravity, including scalar-tensor gravity. This would produce an additional contribution to \pbdot that we can assume to be the residual measurement, 
\begin{equation}
    \dot{P}_{\rm b}^{\rm res} = \dot{P}_{\rm b}^{\rm obs} - \dot{P}_{\rm b}^{\rm kin}. 
\end{equation}
Using the nominal 1$\sigma$ uncertainty of the distance from the L-K corrected estimate of $\varpi$, we obtain  $\dot{P}_{\rm b}^{\rm kin} = 4^{+2}_{-1} \times 10^{-13}$ s~s$^{-1}$ and hence $\dot{P}_{\rm b}^{\rm res} = -2^{+2}_{-3} \times 10^{-13}$  s~s$^{-1}$.

This residual \pbdot can be compared (to leading order and assuming zero contribution from the companion because it is a WD) with the expected \pbdot from $\dot{G}$, 
\begin{equation}
    \dot{P}_{\rm b}^{\dot{G}} = -2\frac{\dot{G}}{G}\left[ 1 - \frac{2M_{\rm p} + 3 M_{\rm c}}{2(M_{\rm p} + M_{\rm c})} s_{\rm p} \right] P_{\rm b},
    \label{eq:pbdot_g}
\end{equation}
where $s_{\rm p}$ is the sensitivity of the NS, which is defined as 
\begin{equation}
    s_{\rm p} \equiv - \frac{\partial M_{\rm p}}{\partial G}\mid_N,
\end{equation}
where N is the fixed number of baryons in the NS \citep{LazaridisEtAl2009}. This sensitivity of an NS depends on the mass, the equation of state (EoS), and the theory of gravity considered. Rewriting Eq.~\eqref{eq:pbdot_g} as 
\begin{equation}
    \frac{\dot{G}}{G} = \frac{\mathbb{A}}{(1-\mathbb{B\,}s_{\rm p})} 
,\end{equation}
we obtain $\mathbb{A} = -1.1^{+1.1}_{-0.7} \times 10^{-19} s^{-1} $ and $\mathbb{B} = 1.115 \pm 0.015$.  Similar to \cite{Zhu2019}, we considered Jordan–Fierz–Brans–Dicke (JFBD) theory and AP4 EoS as an example and find $s_{\rm p} \simeq 0.17$. This provides a limit on $\dot{G}/{G} = -4^{+4}_{-3} \times 10^{-12}$ yr$^{-1}$, consistent with the prediction of  $\dot{G}/{G} = 0$ by General Relativity. Similar tests have been conducted using PSRs J0437$-$4715, J1713+0747 and J1738+0333 \citep{VerbiestEtAl2008, Zhu2019, FreireEtAl2012}, for instance, the most constraining of which is J1713+0747, which is about $\text{four times}$ more sensitive than our results here. Future timing measurements that increase the significance of the timing parallax will aid in performing more stringent tests of $\dot{G}$.

\section{Discussion and conclusions}
\label{sec:discussion}

We have presented the results of our timing of \psrc which combined recent Parkes and MeerKAT timing measurements with earlier Parkes measurements, for a total timing baseline of about 19 years. Because of the high timing precision provided by MeerKAT, the results include precise astrometry, in particular, the first measurement of the parallax of this system, the measurement of several kinematic effects on the binary orbit (including AOP), and a first measurement of its Shapiro delay.
The measurement of the AOP is noteworthy, as this effect has only been detected in four pulsar binaries, namely PSRs J0437$-$4715 (the closest and brightest MSP in the sky, \citep{vanStraten2001}, J2234+0611 \citep{Stovall2019}, J1713+0747 \citep{Zhu2019}, and J2222$-$0137 \citep{Guo2021}.

A detailed analysis of the above effects allowed us for the first time to measure the component masses: $M_{\text{p}} \, = \, 1.4^{+0.3}_{-0.2}$ M$_\odot$ and $M_{\text{c}} = 0.43(5)$ M$_\odot$ and the full orbital orientation of the system ($i = 55(1) \, \deg$, $\Omega = 255^{+8}_{-14} \, \deg$), although the robustness of the latter measurements is seen to depend on the fiducial $T_0$, as seen in Sect.~ \ref{sec:T0dependence}. The root cause of this dependence is currently unclear. An independent measurement of $i$ and $\Omega$ will allow a better understanding of this problem. This independent measurement is possible using scintillation velocity measurements, as has been demonstrated by \cite{ReardonEtAl2018}, although the current data set does not have the necessary frequency resolution needed for the analysis.

Nevertheless, the mass measurements are robust; the companion mass is significantly more massive than the \cite{TaurisAndSavonije1999} prediction for He WDs, indicating that the companion is most likely a CO WD.

We note that the estimated characteristic age of $\sim$24\,Gyr of the pulsar exceeds the Hubble time. This emphasises that for the life cycles of recycled millisecond pulsars, the characteristic age tends to lose its meaning as the underlying assumptions are no longer valid. This implies, for instance, that after recycling, the spin period of this pulsar was close to its current spin period. Nevertheless, we expect this recycled MSP to have a real age of several billion years, such that the WD companion is likely old and cool. Hence, optical observations of \psrc combined with the mass and distance estimates derived herein, can be used to test WD cooling models \citep{Bhalerao:2011ay,Kaplan:2014mka,Bassa:2015qga, Bergeron:2022abc}. Similarly, optical and infrared photometry can constrain the atmospheric composition of the WD, and using the cooling models that survive the tests above, can provide an estimate of its cooling age. These will allow constraining the spin of the pulsar at birth, and placing additional constraints on the accretion history and origin of the system \citep{Bhalerao:2011ay,Tauris:2011ck}. 

By analogy with PSR~J1614$-$2230, it is possible that these fast-spinning pulsars with CO WD companions evolved via case A RLO. The very long accretion episode associated with case A RLO is consistent with the very old characteristic age and low B-field of \psr (see Table~\ref{tab:timing}). Despite this, the
mass of \psr implies that it has not gained more than $\sim 0.2 \, \rm M_{\odot}$. This suggests that accretion is generally extremely inefficient. These conclusions agree with the conclusions of \cite{Tauris:2011ck}, who pointed out that PSR~J1614$-$2230 is massive mainly because it was born this way, with mass transfer accounting for at most $0.2$ - $0.3 \, \msun$. It also agrees with the wider range of MSP masses, where no obvious correlation with spin or orbital parameters has been observed; even the eccentric MSPs, which have a rather uniform set of orbital parameters that suggest a uniform evolutionary mechanism, seem to have a wide range of masses (e.g. \citealt{Serylak2021} and references therein). This provides additional evidence that NS masses are in general acquired at birth, and are not much affected by their subsequent evolution, instead being a product of supernova physics.

Finally, the measurements presented in this work highlight the capabilities of MeerKAT for precise timing and detailed investigations of pulsar binaries. Without the great sensitivity of MeerKAT, most of these results would not have been obtainable. For example, continuing a monthly campaign on \psr for the next five years should lead to an increase in the detection significance of \pbdot by a factor of 3, and consequently, in equal fashion, improve our distance measurements and constraints on $\dot{G}$.
Within the next few years, many other southern binaries will not only have their masses measured accurately, but several of them will also yield new tests of gravity theories from the measurement of multiple PK parameters as part of the MeerTime/RelBin project.

\begin{acknowledgements}
We thank the referee for valuable comments on the manuscript. We thank Norbert Wex, Kuo Liu and Matthew Miles for valuable discussions and Robert Main for comments on the manuscript. The MeerKAT telescope is operated by the South African Radio Astronomy Observatory, which is a facility of the National Research Foundation, an agency of the Department of Science and Innovation. SARAO acknowledges the ongoing advice and calibration of GPS systems by the National Metrology Institute of South Africa (NMISA) and the time space reference systems department department of the Paris Observatory. MeerTime data is housed on the OzSTAR supercomputer at Swinburne University of Technology maintained by the Gravitational Wave Data Centre and ADACS via NCRIS support. The Parkes radio telescope (\textit{Murriyang}) is part of the Australia Telescope National Facility (\href{https://ror.org/05qajvd42}{https://ror.org/05qajvd42}) which is funded by the Australian Government for operation as a National Facility managed by CSIRO. We acknowledge the Wiradjuri people as the traditional owners of the Observatory site. This research has made extensive use of NASA's Astrophysics Data System (https://ui.adsabs.harvard.edu/) and includes archived data obtained through the CSIRO Data Access Portal (http://data.csiro.au). Parts of this research were conducted by the Australian Research Council Centre of Excellence for Gravitational Wave Discovery (OzGrav), through project number CE170100004. VVK, PCCF, MK, JA, MCiB DJC and AP acknowledge continuing valuable support from the Max-Planck Society. JA acknowledges support from the European Commission (Grant Agreement number: 101094354), the Stavros Niarchos Foundation (SNF) and the Hellenic Foundation for Research and Innovation (H.F.R.I.) under the 2nd Call of ``Science and Society -- Action Always strive for excellence -- ``Theodoros Papazoglou'' (Project Number: 01431).
APo and MBu acknowledge the support from the Ministero degli Affari Esteri e della Cooperazione Internazionale - Direzione Generale per la Promozione del Sistema Paese - Progetto di Grande Rilevanza ZA18GR02. MBu and APo acknowledge support through the research grant "iPeska" (PI: Andrea Possenti) funded under the INAF national call Prin-SKA/CTA approved with the Presidential Decree 70/2016. 
RMS acknowledges support through Australian Research Council Future Fellowship FT190100155. J.P.W.V. acknowledges support by the Deutsche Forschungsgemeinschaft (DFG) through the Heisenberg programme (Project No. 433075039).
This publication made use of open source python libraries including Numpy \citep{numpy}, Matplotlib \citep{matplotlib}, Astropy \citep{astropy} and Chain Consumer \citep{Hinton2016_ChainConsumer}, galpy \citep{galpy}, GalDynPsr \citep{Pathak_2018} along with pulsar analysis packages: \textsc{psrchive} \citep{HotanEtAl2004}, \textsc{tempo2} \citep{HobbsEtAl2006}, \textsc{temponest} \citep{LentatiEtAl2014}.
\end{acknowledgements}

\bibliographystyle{aa}
\bibliography{global,main}

\begin{appendix}

\section{Simulations}

\label{sec:app}
We conducted data simulations using the \textsc{toasim} software suite, a plugin to \textsc{tempo2}. This allowed us to test the validity of the obtained orbital inclination ($i$) and longitude of the ascending node ($\Omega$) values as presented in Table \ref{tab:timing2}. We observed that the probabilities associated with ($i$, $\Omega$) and their symmetric solutions ($180^\circ - i$, $360^\circ - \Omega$) depend on T$_0$ , as shown in Fig.~\ref{fig:cosi} and described in Sect. \ref{sec:T0dependence}.

We used the \texttt{formIdeal} function in \textsc{toasim} to obtain idealised ToAs that have the exact same cadence (including the gaps in observing), backends, and jumps as the original data. To do this, we added Gaussian noise (using the \texttt{addGaussian} function) that is statistically equivalent to the noise found to be present in the actual data. To our simulated ToAs we also added stochastic achromatic red noise (using the \texttt{addRedNoise} function) with the same amplitude and spectral index as obtained from the actual data (computed from \textsc{temponest} and presented in Table \ref{tab:timing2}). We executed these simulating steps for two input ephemerides, one ephemeris with its fiducial $T_0$ value at MJD\,53004.13 (simulation 1), and the other with $T_0$ at MJD\,58836.96 (simulation 2). 

After creating these realistic datasets, we ran a \textsc{temponest} 
analysis on each, identical to what was done for real data in Sect.~\ref{sec:timingmodel}, including the same number of live points and other \textsc{multinest} configurations. 
We performed the \textsc{temponest} analysis on both simulations for a range of associated input ephemerides with varying $T_0$ values: MJDs \{53004.13, 55734.66 and 58831.96\}. For simulation 2, we also ran it with two additional input $T_0$ values: MJD\,54260.46 and MJD\,57503.78. This was done to investigate whether we observe any trends in obtained ($i$, $\Omega$) as a function of where we placed $T_0$ and to confirm whether there is any consequence when we place the $T_0$ values at the gaps in the PSR J1933$-$6211 timing baseline. The \textsc{temponest} input ephemerides for both simulated data sets had $i$ and $\Omega$ set to 55.3 and 254 degrees, respectively. As for the real data, the sampling priors on $i$ and $\Omega$ covered all possible values, that is, $0^\circ \leq i \leq 180^\circ$ and $0^\circ  \leq \Omega \leq 360^\circ$.

\begin{figure}
    \label{fig:cosi_sim}
    \centering
    \includegraphics[width=0.45\textwidth]{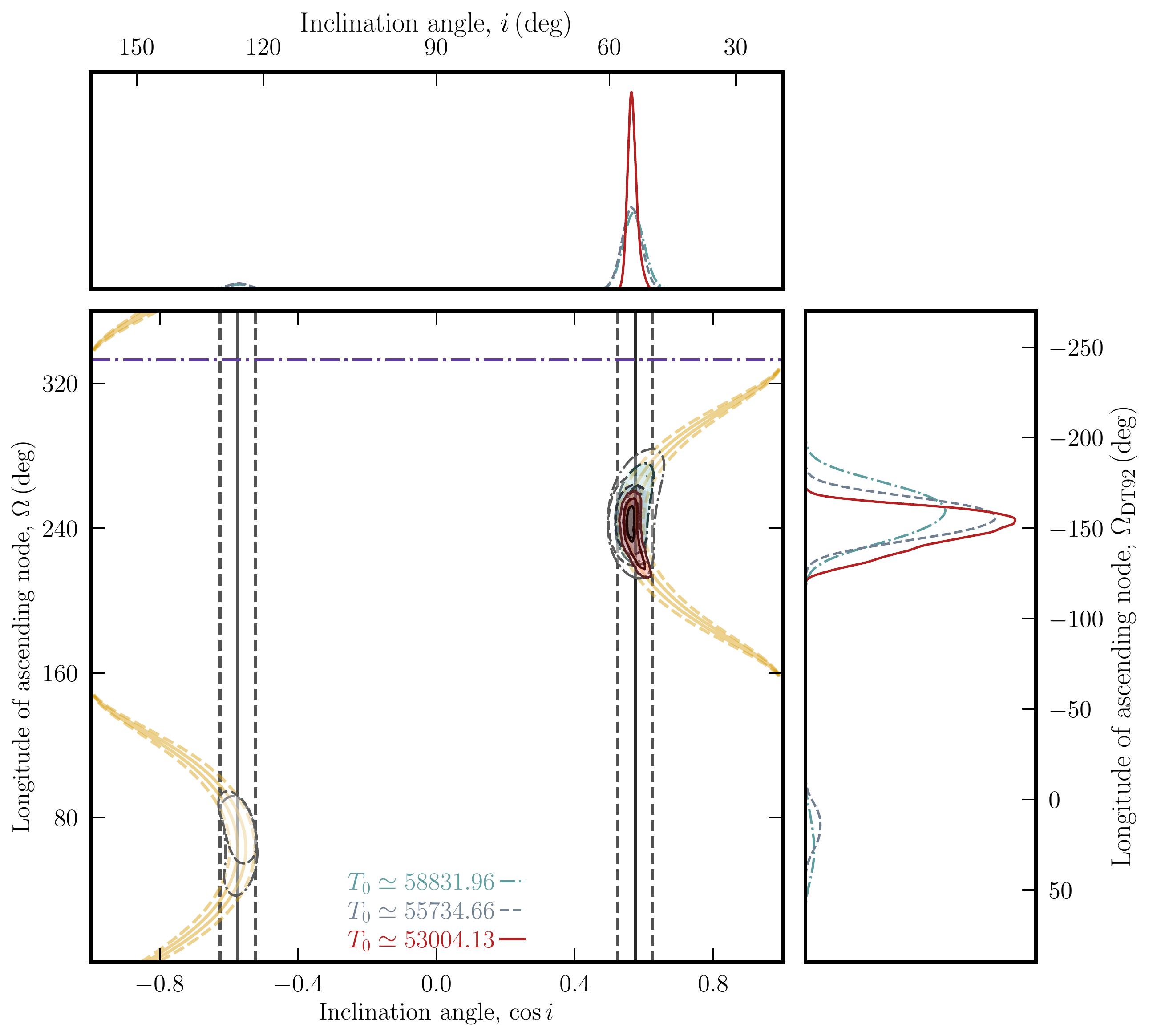}
    
    \includegraphics[width=0.45\textwidth]{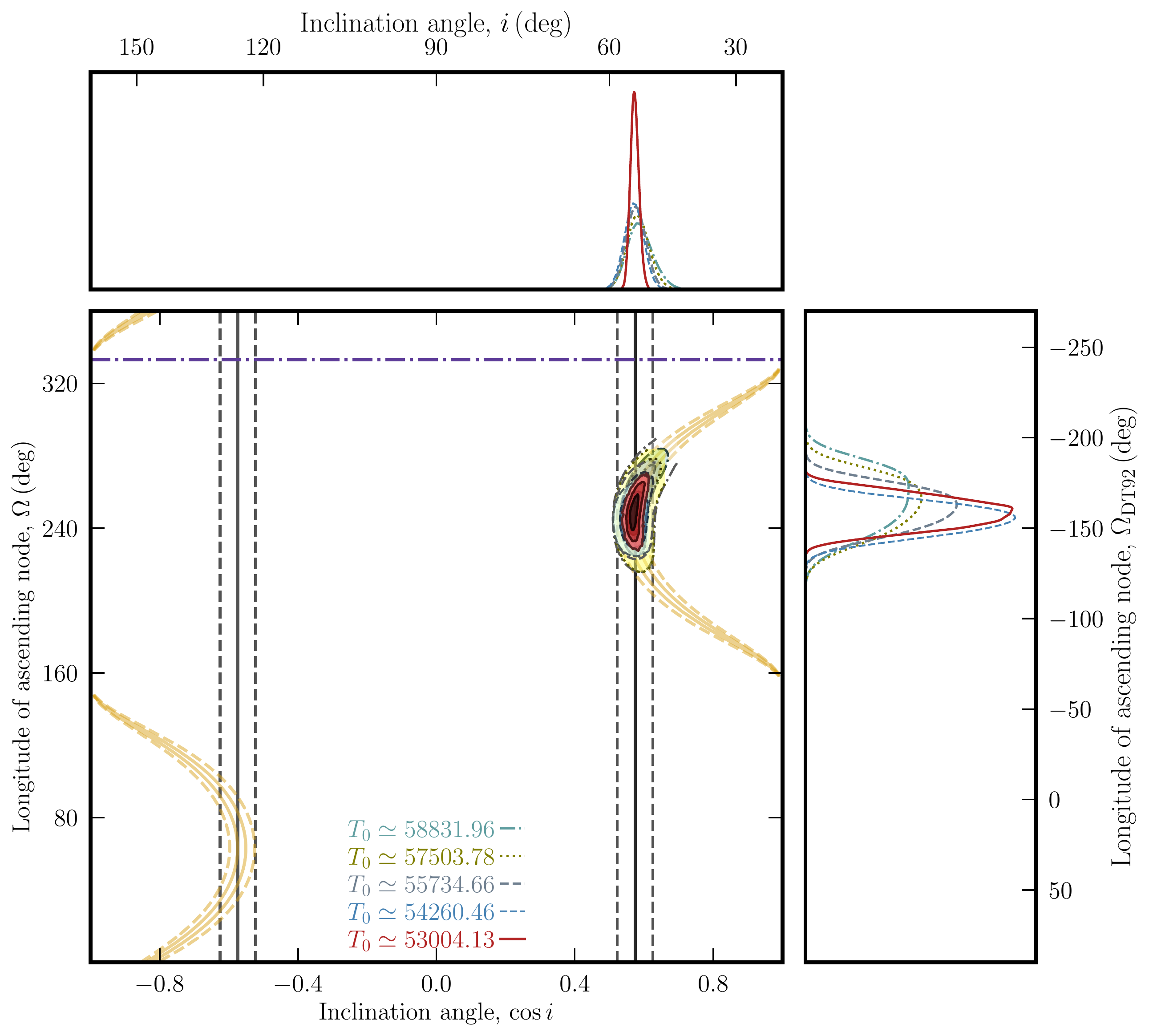}
    \caption{Orbital orientation for the PSR J1933$-$6211 binary as presented in Fig. \ref{fig:cosi}, but using simulated data. In the top panel, the input ephemerides used to generate the fake data set had $T_0$ set to MJD $53004.16$  (simulation 1) and in the bottom panel to MJD $58836.96$ (simulation 2). The \textsc{temponest} analyses were run with varying input ephemerides $T_0$ values, as shown in the plot legends.}
\end{figure}

The results of the simulations are shown in Fig.~\ref{fig:cosi_sim}.1. In none of these simulations do we find significant probabilities for the alternative solution ($i$> 90$^\circ$,  $\Omega$ < 180$^\circ$), in contrast to what we find in the real data. This likely rules out our software as the main reason for this behaviour. With the simulations, we can confirm that for a data set that has the same cadence, noise properties, and residual rms timing precision, we are able to break the degeneracy for the angles $i$ and $\Omega$.

\end{appendix}

\end{document}